%% file: cdw_final.tex
\documentclass[twocolumn,english,prl,superscriptaddress]{revtex4-1}
\usepackage[T1]{fontenc}
\usepackage[latin9]{inputenc}
\usepackage{amsmath}
\usepackage{amssymb}
\usepackage{graphicx}

\makeatletter
\usepackage{babel}

\makeatother

\usepackage{babel}
\begin{document}

\title{Is charge order induced near an antiferromagnetic quantum critical
point?}

\author{Xiaoyu Wang}

\affiliation{School of Physics and Astronomy, University of Minnesota, Minneapolis,
MN 55455}

\author{Yuxuan Wang}

\affiliation{School of Physics and Astronomy, University of Illinois, Urbana-Champaign,
IL 61801}

\author{Yoni Schattner}

\affiliation{Weizmann Institute of Science, Rehovot 7610001, Israel }

\author{Erez Berg}

\affiliation{Department of Physics, University of Chicago, Chicago, IL 60637}

\author{Rafael M. Fernandes}

\affiliation{School of Physics and Astronomy, University of Minnesota, Minneapolis,
MN 55455}
\begin{abstract}
We investigate the interplay between charge order and superconductivity
near an antiferromagnetic quantum critical point using sign-problem-free
Quantum Monte Carlo simulations. We establish that, when the electronic
dispersion is particle-hole symmetric, the system has an emergent
SU(2) symmetry that implies a degeneracy between $d$-wave superconductivity
and charge order with $d$-wave form factor. Deviations from particle-hole
symmetry, however, rapidly lift this degeneracy, despite the fact
that the SU(2) symmetry is preserved at low energies. As a result,
we find a strong suppression of charge order caused by the competing,
leading superconducting instability. Across the antiferromagnetic
phase transition, we also observe a shift in the charge order wave-vector
from diagonal to axial. We discuss the implications of our results
to the universal phase diagram of antiferromagnetic quantum-critical
metals and to the elucidation of the charge order experimentally observed
in the cuprates. 
\end{abstract}
\maketitle
The phase diagrams of a number of strongly correlated materials display
putative quatum critical points (QCP), in which the transition temperature
of an electronically ordered state is suppressed to zero. In systems
displaying antiferromagnetic (AFM) order, such as heavy fermions,
cuprates, and iron pnictides, unconventional superconductivity (SC)
is found to emerge near the QCP \cite{Scalapino12}. Although it is
well established that the interactions mediated by fluctuations near
an AFM-QCP favor a sign-changing SC gap, the extent to which this
physics describes the actual materials remains widely debated. In
this context, analytical investigations of metallic AFM-QCP in two
dimensions revealed a surprising result: the same electronic interaction
that promotes sign-changing SC also promotes an unusual sign-changing
bond charge order (CO) \cite{Metlitski10b,Metlitski10,Efetov13,Yuxuan14,Allais14}.
This magnetic mechanism for CO is sharply distinct from the usual
mechanisms involving phonons and Fermi surface nesting. Taken at face
value, this result would suggest that CO should emerge generically
in the phase diagrams of AFM systems.

These theoretical results were brought to the spotlight by the experimental
observation of sign-changing bond CO in cuprate high-$T_{c}$ superconductors
\cite{Wu11,Chang12,Achkar12,Ghiringhelli12,Blackburn13,Leyraud13,LeBoeuf13,Comin14,Fujita14,Neto14,Mesaros16,Jang16,Chang16},
spurring many ideas on the interplay between AFM-QCP, SC, and CO \cite{Bulut13,Sachdev13,Allais14,Hayward14,Yuxuan14,Grilli17}.
It has been proposed, for instance, that the pseudogap physics is
a manifestation of a more fundamental symmetry between SC and CO near
an AFM-QCP \cite{Efetov13}. However, most of these theoretical works
have relied on certain uncontrolled approximations, which are required
for an analytical treatment of an AFM-QCP in a metal. The fundamental
question about the universality of CO near an AFM-QCP, and the more
specific question about the relevance of this result to explain charge
order in cuprates, beg for unbiased methods to probe this phenomenon.

In this paper, we employ the determinantal Quantum Monte Carlo (QMC)
method to address these questions. We consider the two-band version
of the spin-fermion model, for which the QMC does not suffer from
the fermionic sign-problem \cite{Berg12}. The spin-fermion model~\cite{Abanov03}
consists of free electrons coupled to an AFM order parameter tuned
to its quantum critical point. Analytical and sign-problem-free QMC
calculations have established the existence of a SC dome surrounding
the AFM-QCP in the spin-fermion model \cite{Abanov03,Schattner16}.
As for CO, an emergent SU(2) symmetry between CO and SC was found
analytically within an approximation that considers only the vicinity
of the AFM hot spots \textendash{} the points on the Fermi surface
which are separated by the AFM wave-vector $\mathbf{Q}_{\mathrm{AFM}}=(\pi,\pi)$
\cite{Metlitski10,Efetov13}. The resulting CO wave-vector lies along
the diagonal of the Brillouin zone, $\mathbf{Q}_{\mathrm{CO}}=\left(Q_{0},Q_{0}\right)$,
with $\sqrt{2}Q_{0}$ being the distance between hot spots in momentum
space. CO with axial wave-vectors $\left(Q_{0},0\right)$ and/or $\left(0,Q_{0}\right)$,
which are those experimentally observed in cuprates, has also been
proposed \cite{Yuxuan14,Pepin14,Chowdhury14} within the spin-fermion
model. Although QMC investigations have not yet found CO in the spin-fermion
model, they have focused on very narrow parameter regimes \cite{Schattner16},
or were performed in the superconducting phase \cite{ZiXiang17}.

Here, we report QMC results on the spin-fermion model showing the
existence of an SU(2) symmetry between CO and SC near the AFM-QCP
when the non-interacting band structure has particle-hole symmetry.
This SU(2) symmetry implies a degeneracy between SC and CO, manifested
by a sharp enhancement of both susceptibilities as the AFM-QCP is
approached. This result demonstrates the non-trivial mechanism of
magnetically-mediated CO, and establishes that the same interaction
that promotes SC in the spin-fermion model is also capable of promoting
CO.

As the particle-hole symmetry of the non-interacting band dispersion
is broken, however, we find that while the enhancement of the SC susceptibility
is preserved, the CO susceptibility shows a very weak enhancement
near the QCP. Furthermore, near the onset of SC, the CO susceptibility
is even suppressed with respect to its non-interacting value, signaling
a strong competition between these two states already in the fluctuating
regime. This happens even though the SU(2) symmetry is preserved locally
at the hot spots. The fragility of the CO-SC degeneracy implies that
CO near an AFM-QCP is not a universal phenomenon, but instead requires
a fine-tuned band structure that goes beyond just hot-spot properties.
We also investigate the wave-vector for which the CO susceptibility
is maximal. When CO and SC are degenerate, the wave-vector is diagonal,
in agreement with the analytical approximations. However, once CO
and SC are no longer degenerate, the wave-vector tends to change from
diagonal to axial as the AFM-QCP is approached. This is consistent
with theoretical proposals that axial CO is favored over the diagonal
one if the anti-nodal region of the Brillouin zone is gapped \cite{Chowdhury14b,Atkinson15}.
Finally, we discuss the implications of our results to materials that
display putative AFM-QCPs and their relevance to understand CO in
the cuprates.

The spin-fermion model is a low-energy model describing electrons
interacting via the exchange of AFM fluctuations. In its two-band
version (whose physics has been argued to be similar to the one-band
version \cite{Xiaoyu17}), the model is described by the following
action, $S=S_{\psi}+S_{\phi}+S_{\lambda}$, defined on a two-dimensional
square lattice: 
\begin{align}
S_{\psi} & =\int_{\tau,\mathbf{r}\mathbf{r}^{\prime}}\sum_{i=c,d}\left[\left(\partial_{\tau}-\mu\right)\delta_{\mathbf{r}\mathbf{r}^{\prime}}-t_{i,\mathbf{r}\mathbf{r}^{\prime}}\right]\psi_{i,\mathbf{r}\alpha}^{\dagger}\psi_{i,\mathbf{r}^{\prime}\alpha}\nonumber \\
S_{\phi} & =\frac{1}{2}\int_{\tau,\mathbf{r}}\left[\frac{1}{v_{s}^{2}}\left(\partial_{\tau}\boldsymbol{\phi}\right)^{2}+\left(\nabla\boldsymbol{\phi}\right)^{2}+r_{0}\boldsymbol{\phi}^{2}+\frac{u}{2}\left(\boldsymbol{\phi}^{2}\right)^{2}\right]\nonumber \\
S_{\lambda} & =\lambda\int_{\tau,\mathbf{r}}e^{i\mathbf{Q}_{\mathrm{AFM}}\cdot\mathbf{r}}\boldsymbol{\phi}\cdot\left(\psi_{c,\mathbf{r}\alpha}^{\dagger}\boldsymbol{\sigma}_{\alpha\beta}\psi_{d,\mathbf{r}\beta}+h.c.\right)\label{action}
\end{align}

Here, $\int_{\tau,\mathbf{r}}$ is shorthand for $\int d\tau\sum_{\mathbf{r}}$,
$\tau\in[0,\beta)$ is the imaginary time, and $\beta=1/T$ is the
inverse temperature. The action $S_{\psi}$ describes the fermionic
degrees of freedom, with the operator $\psi_{i,\mathbf{r}\alpha}$
annihilating an electron of spin $\alpha$ at site $\mathbf{r}$ and
band $i$. Summation over $\alpha,\beta$ is implied. There are two
different bands, labeled $c$ and $d$. The band dispersion is parametrized
by the chemical potential $\mu$ and the hopping amplitudes $t_{i,\mathbf{r}\mathbf{r}^{\prime}}$.
Here, we consider only nearest-neighbor hopping and set $t_{c,x}=t_{d,y}\equiv t_{x}$
and $t_{c,y}=t_{d,x}\equiv t_{y}$ to enforce the system to remain
invariant under a $90^{\circ}$ rotation followed by a $c\leftrightarrow d$
exchange. The action $S_{\phi}$ describes the spin degrees of freedom,
with the bosonic field $\boldsymbol{\phi}$ denoting the antiferromagnetic
order parameter with ordering wave-vector $\mathbf{Q}_{\mathrm{AFM}}=(\pi,\pi)$,
and $\boldsymbol{\sigma}$ denoting Pauli matrices. The parameter
$r_{0}$ tunes the AFM transition to $T=0$ at $r_{0}=r_{c}$, whereas
$v_{s}$ and $u$ describe the stiffness of AFM temporal and amplitude
fluctuations, respectively. To save computational time, we follow
previous works and consider easy-plane antiferromagnetism, i.e $\boldsymbol{\phi}=(\phi_{x},\phi_{y})$
\cite{Schattner16,Gerlach17,Xiaoyu17}. The action $S_{\lambda}$
couples spins and fermions via the parameter $\lambda$. The two-band
structure of the model ensures the absence of the sign problem in
our simulations \cite{Berg12}.

The fermionic, magnetic, and superconducting properties of this model
have been thoroughly studied recently, revealing a SC dome surrounding
the QCP \cite{Schattner16,Gerlach17}. In particular, the SC order
parameter $\Delta$ was found to have a ``$d$-wave'' symmetry,
i.e. to change its sign between the two bands: $\Delta=\int_{\tau,\mathbf{r}}i\sigma_{\alpha\beta}^{y}\left(\psi_{c,\mathbf{r}\alpha}\psi_{c,\mathbf{r}\beta}-\psi_{d,\mathbf{r}\alpha}\psi_{d,\mathbf{r}\beta}\right).$
The CO order parameter $\rho$ investigated here also has opposite
signs in the two-bands (and is thus analogous to the $d$-wave bond
CO in the one-band version of the model):
\begin{align}
\rho=\int_{\tau,\mathbf{r}}e^{i\mathbf{Q}_{\mathrm{CO}}\cdot\mathbf{r}}\sigma_{\alpha\beta}^{0}\left(\psi_{c,\mathbf{r}\alpha}^{\dagger}\psi_{c,\mathbf{r}\alpha}-\psi_{d,\mathbf{r}\alpha}^{\dagger}\psi_{d,\mathbf{r}\beta}\right).
\end{align}

\begin{figure}
\begin{centering}
\includegraphics[width=0.8\linewidth]{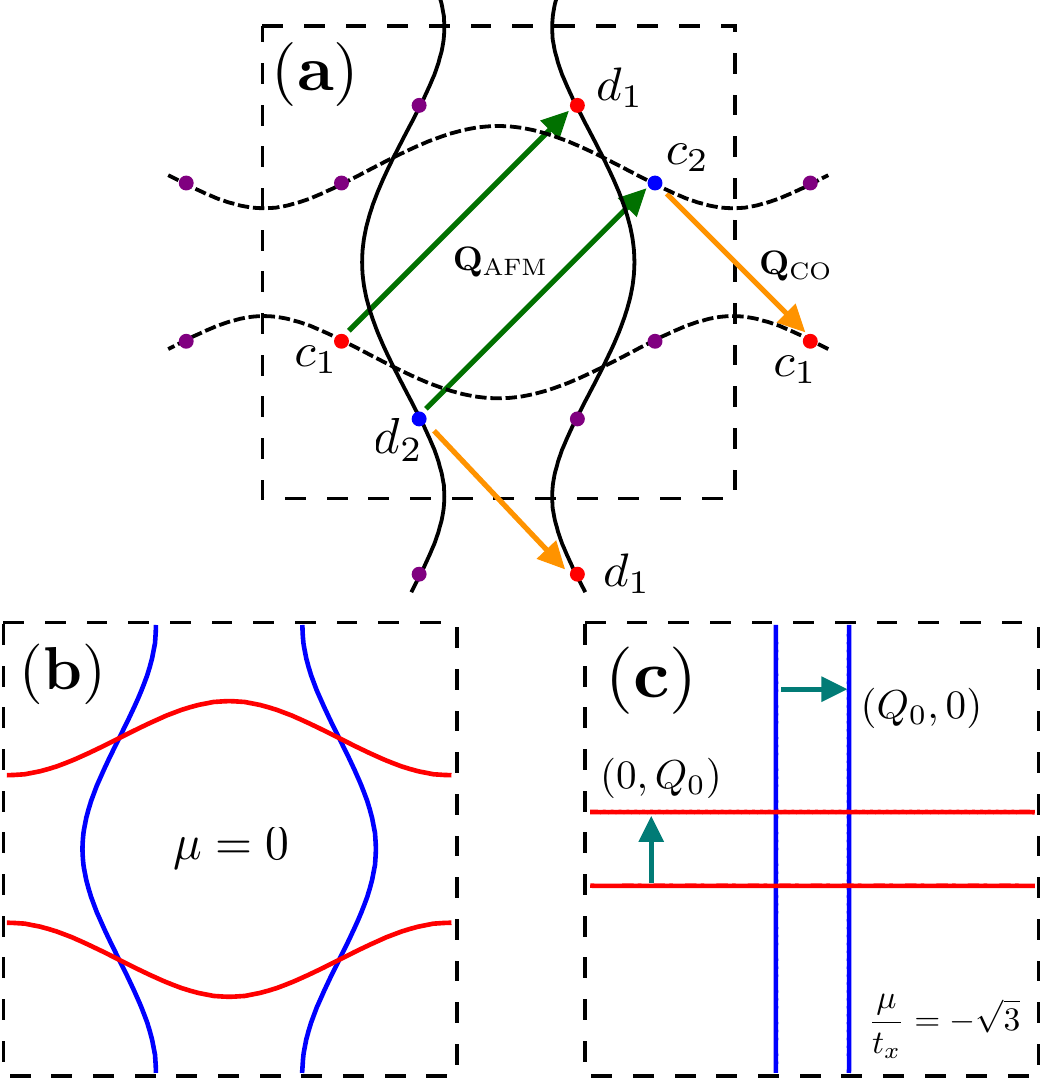} 
\par\end{centering}
\caption{\label{fig:dispersions}(a) Schematic Fermi surface of the spin-fermion
model with two bands ($c$, dashed line, and $d$, solid line). Hot
spots are marked by solid symbols. Two pairs of hot spots $(c_{1},d_{1})$
and $(c_{2},d_{2})$ are highlighted to illustrate the relationship
between the AFM wave-vector $\mathbf{Q}_{\text{AFM}}$ and the CO
wave-vector $\mathbf{Q}_{\text{CO}}$. The band dispersions used in
our QMC calculations are shown in (b) (particle-hole symmetric dispersion,
$\mu=0$, with $t_{y}=t_{x}/2$) and (c) (particle-hole asymmetric
dispersion, $\mu/t_{x}=-\sqrt{3}$, with $t_{y}=0$). Changing $\mu$
tunes the CO wave-vector $\mathbf{Q}_{\mathrm{CO}}=(Q_{\mathrm{0}},\,Q_{0})$
since $Q_{0}=2\arccos\frac{-\mu}{2t_{x}}$.}
\end{figure}
where $\mathbf{Q}_{\mathrm{CO}}$ is the CO wave-vector. Analytical
studies of the spin-fermion model found a special symmetry relating
the SC and CO order parameters under an approximation that focuses
on the hot spots of the model, i.e. the Fermi surface points separated
by $\mathbf{Q}_{\mathrm{AFM}}=(\pi,\pi)$ \cite{Metlitski10,Efetov13,Yuxuan14}.
In the two-band version of the model, each hot spot of a given pair
$(c_{i},d_{i})$ is located on a different band, as shown in Fig.\ \ref{fig:dispersions}.
According to \cite{Metlitski10,Efetov13,Yuxuan15}, the hot-spots
model with linearized dispersions has an emergent symmetry that rotates
the SC order parameter, $\Delta=i\sigma_{\alpha\beta}^{y}\left(\psi_{c_{1},\alpha}\psi_{c_{2},\beta}-\psi_{d_{1},\alpha}\psi_{d_{2},\beta}\right)$,
onto the CO order parameter, $\rho=\sigma_{\alpha\beta}^{0}\left(\psi_{c_{1},\alpha}^{\dagger}\psi_{c_{2},\beta}-\psi_{d_{1},\alpha}^{\dagger}\psi_{d_{2},\beta}\right)$.
Note that this CO has a diagonal wave-vector $\mathbf{Q}_{\mathrm{CO}}\equiv(Q_{\mathrm{0}},\,Q_{0})$
which separates two hot spots belonging to different pairs but to
the same band (see Fig. \ref{fig:dispersions}). Our goal here is
to investigate: (i) to what extent does this symmetry play a role
in the vicinity of an AFM-QCP, and (ii) more broadly, is CO a generic
feature near such a QCP. To this end, we perform a systematic investigation
of the SC and CO susceptibilities in the two-band spin-fermion model. 

We choose as our starting point the parameters for which the symmetry
of the low-energy hot-spots model discussed above is promoted to an
exact lattice symmetry. This corresponds to the case where the $c$
and $d$ bands are particle-hole symmetric, i.e. $\mu=0$. This allows
us to systematically study the effect of breaking the particle-hole
symmetry at the lattice level. For $\mu=0$, the electronic action
for a given AFM field configuration \textendash{} corresponding to
the $S_{\psi}$ and $S_{\lambda}$ terms of the action in Eq. (\ref{action})
\cite{Metlitski10}\textendash{} is invariant under a rotation in
particle-hole space, $\psi_{i,\mathbf{r}\alpha}\rightarrow e^{i\mathbf{Q}_{\mathrm{AFM}}\cdot\mathbf{r}}\left(i\sigma_{\alpha\beta}^{y}\right)\psi_{i,\mathbf{r}\beta}^{\dagger}$.
This invariance can be seen by constructing a four-dimensional spinor
that combines rotated and non-rotated operators at each band, $\Psi_{i,\mathbf{r}}\equiv\left(\psi_{i,\mathbf{r}\uparrow},\ \psi_{i,\mathbf{r}\downarrow},\ e^{i\mathbf{Q}_{\mathrm{AFM}}\cdot\mathbf{r}}\psi_{i,\mathbf{r}\downarrow}^{\dagger},\ -e^{i\mathbf{Q}_{\mathrm{AFM}}\cdot\mathbf{r}}\psi_{i,\mathbf{r}\uparrow}^{\dagger}\right)^{T}$.
In this representation, when $\mu=0$, the Hamiltonian commutes with
all all SU(2) generators $\boldsymbol{\tau}$ in particle-hole space.
Importantly, the SC and CO order parameters form a three-component
vector $\boldsymbol{\Phi}\equiv\left(\mathfrak{Re}\Delta,\mathfrak{Im}\Delta,\rho\right)$
in this space, which couples to the electrons as $\sum_{\mathbf{r}}e^{i\mathbf{Q}_{\mathrm{AFM}}\cdot\mathbf{r}}\boldsymbol{\Phi}\cdot\left(\sigma_{0}\otimes\boldsymbol{\tau}\right)\left(\Psi_{c,\mathbf{r}}^{\dagger}\Psi_{c,\mathbf{r}}-\Psi_{d,\mathbf{r}}^{\dagger}\Psi_{d,\mathbf{r}}\right)$.
Note that $\mathbf{Q}_{\mathrm{CO}}=\mathbf{Q}_{\mathrm{AFM}}=(\pi,\pi)$,
enforcing $\rho$ to be real. As a result, an enhancement of the SC
susceptibility also implies an equally strong enhacement in the CO
channel, since the two order parameters are related by rotations in
the SU(2) particle-hole space, and the Hamiltonian is invariant under
these rotations. This symmetry is analogous to the degeneracy between
SC and CO observed in the half-filled negative-$U$ Hubbard model
\cite{Moreo91}. Here, however, both the SC and CO have a $d$-wave
symmetry. 

\begin{figure}
\centering{}\includegraphics[width=0.95\linewidth]{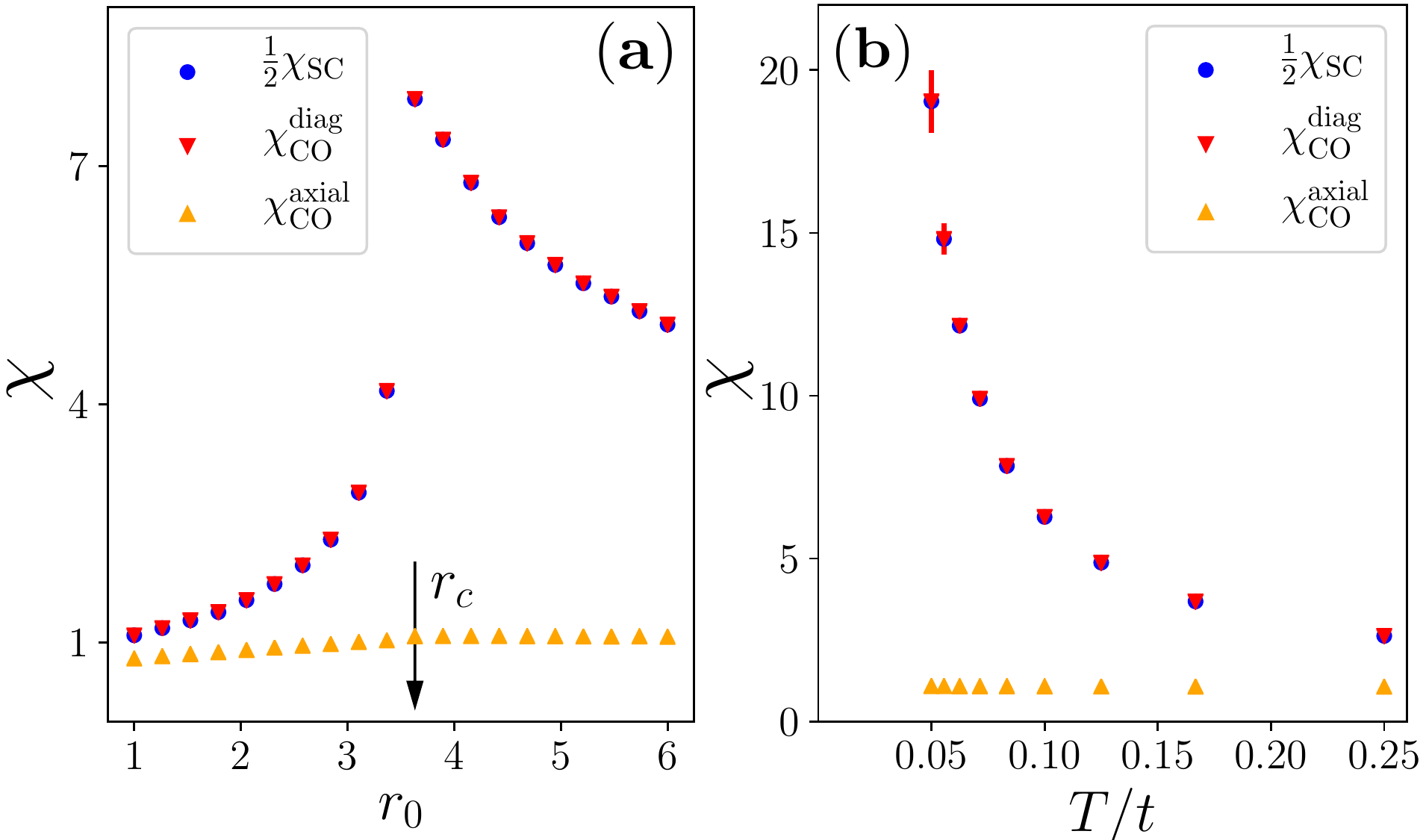}\caption{\label{fig:hf_numerics}SC susceptibility $\chi_{\mathrm{SC}}$ (circles)
and CO susceptibilities for diagonal wave-vector $\mathbf{Q}_{\mathrm{CO}}=\left(Q_{0},Q_{0}\right)$,
$\chi_{\mathrm{CO}}^{\mathrm{diag}}$ (triangles), and axial wave-vector
$\mathbf{Q}_{\mathrm{CO}}=\left(Q_{0},0\right)/\left(0,Q_{0}\right)$,
$\chi_{\mathrm{CO}}^{\mathrm{axial}}$ (inverted triangles), as function
of: (a) the distance $r_{0}-r_{c}$ to the AFM-QCP (fixed temperature
$\beta t=12$); and (b) temperature $T/t$ (fixed $r_{0}=r_{c}$ at
the AFM-QCP). The particle-hole symmetric dispersion used here is
that of Fig. \ref{fig:dispersions}(b).}
\end{figure}

To demonstrate the existence of this SU(2) symmetry for $\mu=0$,
we perform QMC simulations on a square lattice of size $L=12$. Additional
details of the QMC procedure can be found elsewhere \cite{Schattner16}.
All energies are expressed in terms of the hopping $t_{x}\equiv t$
and the parameters are set to $v_{s}=2t$, $u=t^{-1}$, $\lambda^{2}=4t$,
and $t_{y}=t/2$, resulting in the Fermi surface illustrated in Fig.
\ref{fig:dispersions}(b) (the results are the same for other values
of $t_{y}$, see Supplementary Material). Fig. \ref{fig:hf_numerics}(a)
shows the SC susceptibility $\chi_{\mathrm{SC}}$, the CO susceptibility
$\chi_{\mathrm{CO}}^{\mathrm{diag}}$ with diagonal wave-vector $\mathbf{Q}_{\mathrm{CO}}=\left(Q_{0},Q_{0}\right)$,
where $Q_{0}=\pi$, and the CO susceptibility $\chi_{\mathrm{CO}}^{\mathrm{axial}}$
with axial wave-vector $\mathbf{Q}_{\mathrm{CO}}=\left(Q_{0},0\right)/\left(0,Q_{0}\right)$
as a function of the distance to the AFM-QCP for $\beta t=12$. The
position $r_{c}$ of the AFM-QCP was determined via the AFM susceptibility
\cite{Xiaoyu17}. The degeneracy between diagonal CO and SC is evident,
as well as the enhancement of both susceptibilities at the AFM-QCP.
The fact that $\chi_{\mathrm{SC}}=2\chi_{\mathrm{CO}}^{\mathrm{diag}}$
is because the complex SC order parameter has two components whereas
the real CO order parameter has one. In contrast, the axial CO susceptibility
remains small and nearly unaffected by the proximity to the QCP. Fig.
\ref{fig:hf_numerics}(b), which shows the behavior at the QCP, confirms
that the degeneracy is present at all temperatures.

\begin{figure}
\begin{centering}
\includegraphics[width=0.95\linewidth]{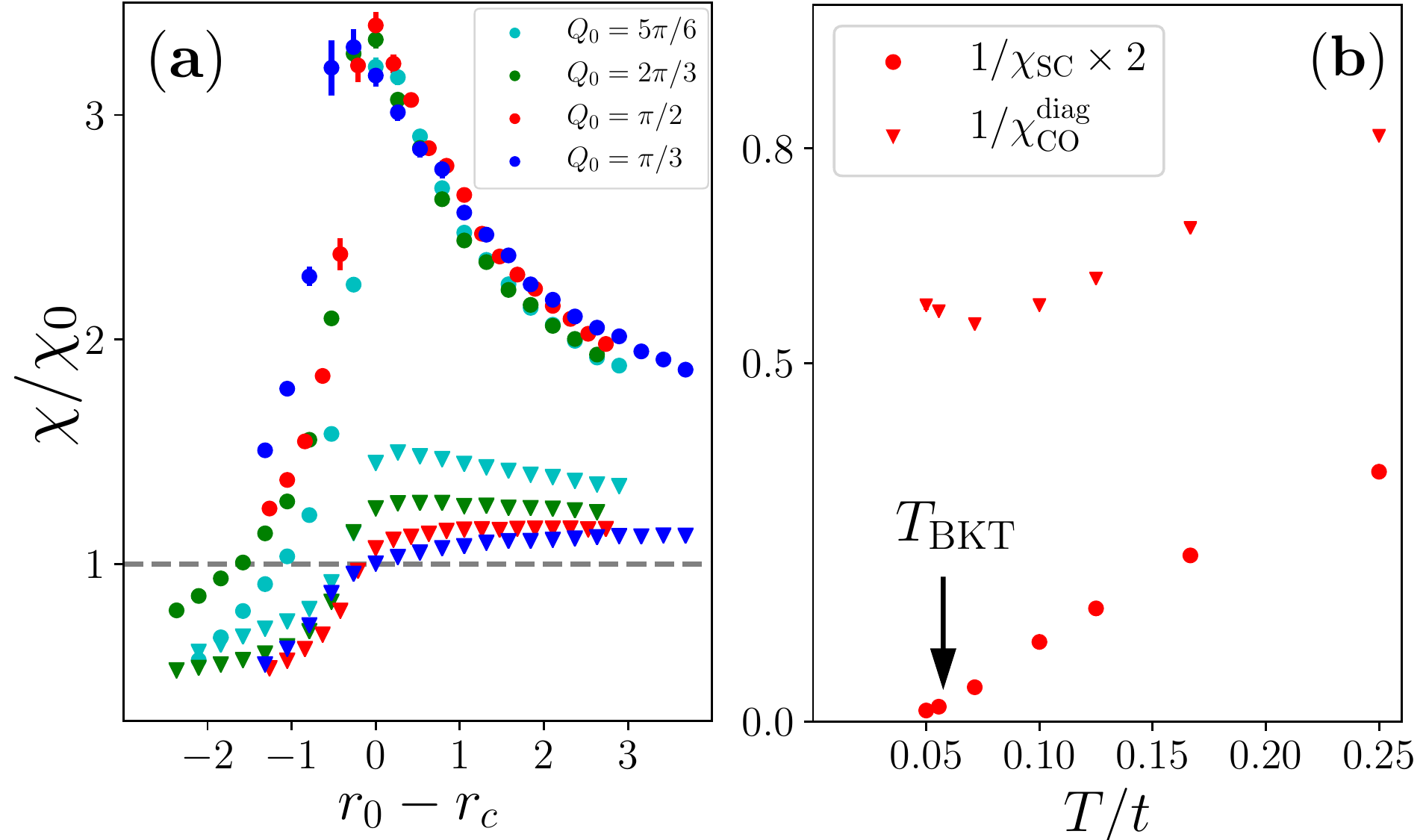} 
\par\end{centering}
\caption{\label{fig:sus_band}(a) SC (circles) and diagonal CO (triangles)
susceptibilities, normalized by their non-interacting values, as a
function of the distance to the QCP $r_{0}-r_{c}$ and for a fixed
temperature $\beta t=10$. The dispersion is represented in Fig. \ref{fig:dispersions}(c),
with different values of the wave-vector $Q_{0}$ (shown in the inset).
Panel (b) shows the temperature dependence of the inverse susceptibilities
at the AFM-QCP ($r_{0}=r_{c}$) for $\mu/t=-\sqrt{2}$ ($Q_{0}=\pi/2$).}
\end{figure}

We now proceed to investigate whether there is a remnant near-degeneracy
between SC and CO when particle-hole symmetry is broken ($\mu\neq0$).
In this case, although there is no lattice SU(2) symmetry, an approximate
SU(2) symmetry of the low energy theory near the hot spots is preserved
\cite{Yuxuan15}\footnote{Technically the hot-spots symmetry is SU(2)$\times$ SU(2)$\sim$SO(4)
and not SU(2), because for $\mu\neq0$ the CO order parameter is complex,
giving rise to a four-component super-vector.}. To favor the CO state, we consider one-dimensional dispersions ($t_{y}=0$),
as shown in Fig. \ref{fig:dispersions}(c), although the results are
similar for finite $t_{y}$ (see Supplementary Material). To be able
to assess the relevant CO wave-vectors $\mathbf{Q}_{\mathrm{CO}}=\left(Q_{0},Q_{0}\right)$
in the finite-size QMC simulations, we choose $\mu$ values that yield
commensurate $Q_{0}\equiv2\arccos\frac{-\mu}{2t}=\frac{2\pi n}{L}$,
namely: $\mu/t=-2\cos\frac{5\pi}{12}\approx-0.52$ ($Q_{0}=5\pi/6$),
$\mu/t=-1$ ($Q_{0}=2\pi/3$), $\mu/t=-\sqrt{2}$ ($Q_{0}=\pi/2$),
and $\mu/t=-\sqrt{3}$ ($Q_{0}=\pi/3$).

Figure \ref{fig:sus_band}(a) displays the behavior of $\chi_{\mathrm{SC}}$
and $\chi_{\mathrm{CO}}^{\mathrm{diag}}$, normalized by their non-interacting
($\lambda=0$) value, as a function of the distance to the QCP for
different values of $\mu$. While the sharp enhancement of $\chi_{\mathrm{SC}}$
at $r_{0}=r_{c}$ is preserved, the enhancement of $\chi_{\mathrm{CO}}^{\mathrm{diag}}$
is small for $r_{0}>r_{c}$. This enhancement of $\chi_{\mathrm{CO}}^{\mathrm{diag}}$
is larger the closer $\mu$ is to zero, i.e. the closer the global
lattice SU(2) symmetry is to be restored. The CO-SC degeneracy observed
for $\mu=0$ is thus removed, with SC clearly winning over CO. The
competition between the two orders is highlighted in Fig.\ \ref{fig:sus_band}(b),
where the $T$ dependences of $1/\chi_{\mathrm{SC}}$ and $1/\chi_{\mathrm{CO}}^{\mathrm{diag}}$
are plotted at the QCP, $r_{0}=r_{c}$. Interestingly, right above
the Berezinskii-Kosterlitz-Thouless superconducting transition temperature
(extracted from the superfluid density,\textbf{ }see Ref. \cite{Xiaoyu17}),
$\chi_{\mathrm{CO}}^{\mathrm{diag}}$ reverses its trend and starts
to decrease upon lowering the temperature. This provides evidence
that the competition between SC and CO takes place already in the
fluctuating regime.

\begin{figure}
\begin{centering}
\includegraphics[width=0.95\linewidth]{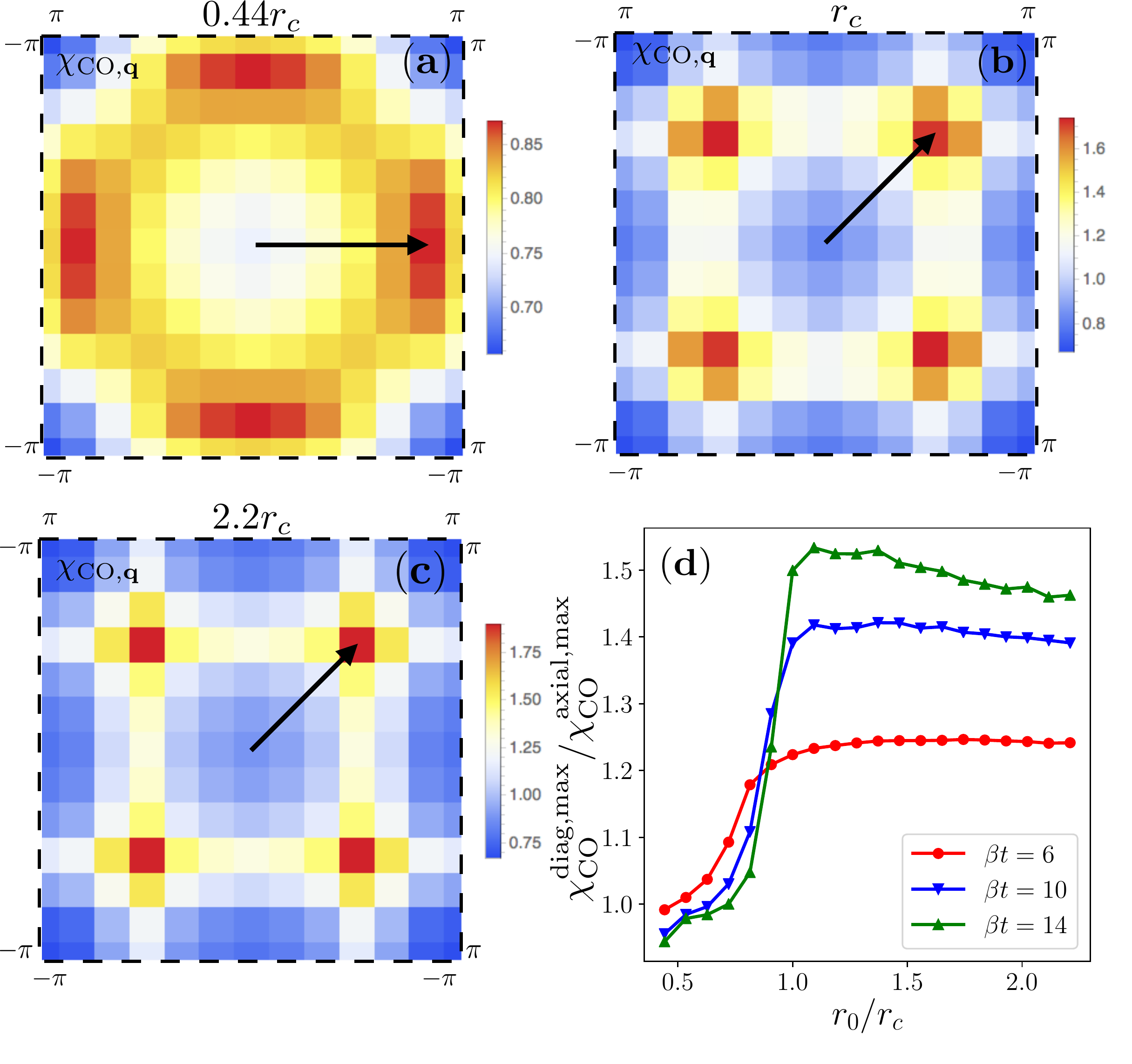} 
\par\end{centering}
\caption{\label{fig:cdw_wavevector}Panels (a)-(c) show the momentum dependence
of the CO susceptibility $\chi_{\mathrm{CO}}\left(\mathbf{q}\right)$
in the AFM phase (a), at the QCP (b), and in the disordered phase
(c) for $\beta t=14$. The dispersion is represented in Fig. \ref{fig:dispersions}(c)
with $\mu/t=-\sqrt{2}$. (b) Ratio of the maximum values of $\chi_{\mathrm{CO}}\left(\mathbf{q}\right)$
along the diagonal direction, $\mathbf{q}=(q,q)$, and along the axial
direction, $\mathbf{q}=(q,0)$, for different inverse temperatures
$\beta$ (inset), as function of the distance to the QCP at $r_{0}=r_{c}$.}
\end{figure}

Another important result from our QMC simulations is that, when the
AFM hot spots are near the antinodal region of the Brillouin zone
(i.e., $\left(\pi,0\right)/\left(0,\pi\right)$), the CO wave-vector
tends to shift from diagonal to axial inside the AFM phase. To illustrate
this, in Fig.\ \ref{fig:cdw_wavevector} we plot $\chi_{\mathrm{CO}}\left(\mathbf{q}\right)$
for the system with $\mu/t=-\sqrt{2}$ ($Q_{0}=\pi/2$) in the disordered
phase ($r_{0}>r_{c}$), at the QCP ($r_{0}=r_{c}$), and in the AFM
phase ($r_{0}<r_{c}$). Results for other fillings are discussed in
the Supplementary Material. The tendency of shifting $\mathbf{Q}_{\mathrm{CO}}$
from diagonal (above the QCP) to axial (below the QCP) is evident.
To quantify this behavior, we plot in Fig. \ref{fig:cdw_wavevector}(d)
the ratio between the maxima of $\chi_{{\rm CO}}$ along the diagonal
and axial directions as function of $r_{0}$ for different temperatures.
Upon approaching the QCP from the disordered side, and upon decreasing
the temperature, this ratio increases due to the enhancement of diagonal
CO by quantum critical AFM fluctuations. Below the QCP and inside
the AFM phase ($r_{0}<r_{c}$), however, the maximum of $\chi_{\text{CO}}(\mathbf{q})$
quickly shifts to the axial direction, and the temperature dependence
of the ratio is the opposite as in the disordered side. This effect
is consistent with theoretical proposals that axial CO is favored
over diagonal CO if the antinodal regions of the Brilliouin zone are
gapped (e.g.\ by AFM order here or by a more exotic pseudogap state
\cite{Chowdhury14b,Atkinson15}).

In summary, we showed that the spin-fermion model with particle-hole
symmetric bands has a global SU(2) symmetry that relates $d$-wave
SC and $d$-wave CO. The breaking of this particle-hole symmetry strongly
suppresses the CO susceptibility, even though the SU(2) symmetry is
still present near the hot spots. Compared with previous QMC investigations
of the spin-fermion model, which showed that the SC instability is
governed by the hot spots \cite{Xiaoyu17}, our results indicate that
the CO instability is instead governed by the full electronic dispersion.
Such an asymmetry between CO and SC implies that CO is not a universal
phenomenon associated with AFM quantum criticality, in contrast to
SC. 

The applicability of these results to specific materials \textendash{}
and particularly the cuprates \textendash{} remains an open question.
On the one hand, the CO observed in most cuprates only acquires a
substantially long correlation length once SC is fully suppressed,
and CO fluctuations are found to be suppressed by the onset of SC
\cite{Chang12,Chang16,Jang16}. Furthermore, in the pseudogap state
where CO is experimentally observed, the CO wave-vector is axial,
and not diagonal. All these observations seem at least qualitatively
consistent with our results for systems without particle-hole symmetric
band dispersions. On the other hand, in hole-doped cuprates, AFM fluctuations
become weaker as the system approaches optimal doping and CO is observed.
The fact that CO is strongest near a specific doping close to $1/8$,
where AFM fluctuations are not particularly enhanced, suggests that
lattice commensuration effects, rather than AFM criticality, may play
an important role in these systems.
\begin{acknowledgments}
We thank A. Chubukov for fruitful discussions. X.W. and R.M.F. were
supported by the U.S. Department of Energy, Office of Science, Basic
Energy Sciences, under Award number DE-SC0012336. R.M.F. also acknowledges
partial support from the Research Corporation for Science Advancement
via the Cottrell Scholar Award, and X.W. acknowledges support from
the Doctoral Dissertation Fellowship offered by the University of
Minnesota. Y.W. is supported by the Gordon and Betty Moore Foundation\textquoteright s
EPiQS Initiative through Grant No. GBMF4305 at the University of Illinois.
R.M.F. and X.W. thank the Minnesota Supercomputing Institute (MSI)
at the University of Minnesota, where part of the numerical computations
was performed. 
\end{acknowledgments}

\bibliographystyle{apsrev4-1}
\input{cdw_final.bbl}

\pagebreak
\widetext
\begin{center}
\textbf{\large Supplementary Material: ``Is charge order induced near an antiferromagnetic
quantum critical point?''}
\end{center}
\setcounter{equation}{0}
\setcounter{figure}{0}
\setcounter{table}{0}
\setcounter{page}{1}
\renewcommand{\theequation}{S\arabic{equation}}
\renewcommand{\thefigure}{S\arabic{figure}}
\renewcommand{\bibnumfmt}[1]{[S#1]}
\makeatother

\section{Degeneracy between SC and CO at half-filling }

\begin{figure}
\includegraphics[width=0.95\linewidth]{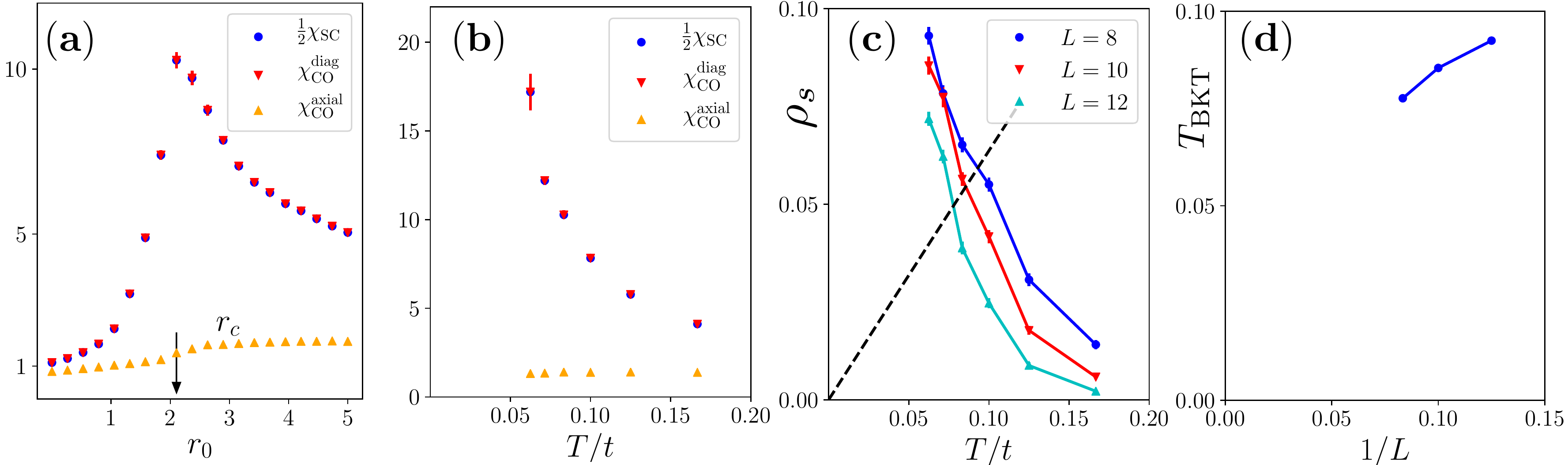}\caption{\label{fig:hf}QMC results for the one-dimensional electronic band
dispersion ($t_{y}=0$) at half-filling $\mu=0$. (a) Sign-changing
SC and CO susceptibilities as a function of the distance to AFM-QCP.
Results obtained for $\beta t=12$ (b) Temperature evolution of SC
and CO susceptibilities at the AFM-QCP. Both (a) and (b) are obtained
for the system size $L=12$. (c) Temperature evolution of the superfluid
density at the AFM-QCP, extracted from the current-current correlation
function, see Ref. \cite{Schattner16,Xiaoyu17}. Different system
sizes are plotted. The shift of the AFM-QCP with system size is negligible.
(d) Berezinskii-Kosterlitz-Thouless transition temperature as a function
of inverse system size, extracted from the superfluid density. }

\end{figure}

When the two-band spin-fermion model has the exact lattice symmetry
(half-filling), the $d$-wave superconductivity (SC) and $d$-wave
charge order (CO) with wave-vector $\mathbf{Q}_{\text{CO}}=\left(\pi,\pi\right)$
form a three-component super-vector. At any temperature and distance
to the antiferromagnetic quantum critical point (AFM-QCP), the CO
and SC susceptibilities differ by a factor of 2, corresponding to
having a real CO and a complex SC order parameter. While in the main
text we focused on the particular band dispersion with $t_{y}=t_{x}/2$,
the CO-SC degeneracy holds for any band dispersion at half-filling
($\mu=0$). To verify this, we also considered the purely one-dimensional
band dispersion at half-filling, with $t_{y}=0$. In Figure \ref{fig:hf},
we present the SC and CO susceptibilities as a function of the distance
to the AFM-QCP for the inverse temperature $\beta t=12$, as well
as at various temperatures above the QCP. The SC susceptibility is
rescaled by $\frac{1}{2}$. The degeneracy between SC and CO is evident.

The enlarged SU(2) symmetry also means that there is no finite-temperature
Berezinskii-Kosterlitz-Thouless phase transition. In Fig. \ref{fig:hf},
we show the superfluid density as well as the extracted BKT transition
temperature for system sizes $L=8$, $L=10$, and $L=12$. As $L$
increases, the superfluid density at a given temperature decreases,
and that the BKT temperature shows finite size scaling. Due to computational
costs, we did not go to larger system sizes. Nonetheless, the fact
that $T_{\text{BKT}}$ goes down with system size is consistent with
having an enlarged symmetry. For comparison, in systems without the
enlarged symmetry, $T_{\mathrm{BKT}}$ remains nearly saturated for
these system sizes (see Ref. \cite{Xiaoyu17}).

\section{Evolution of the CO wave-vector away from half-filling}

\subsection{1D dispersions}

\begin{figure}
\includegraphics[width=0.95\linewidth]{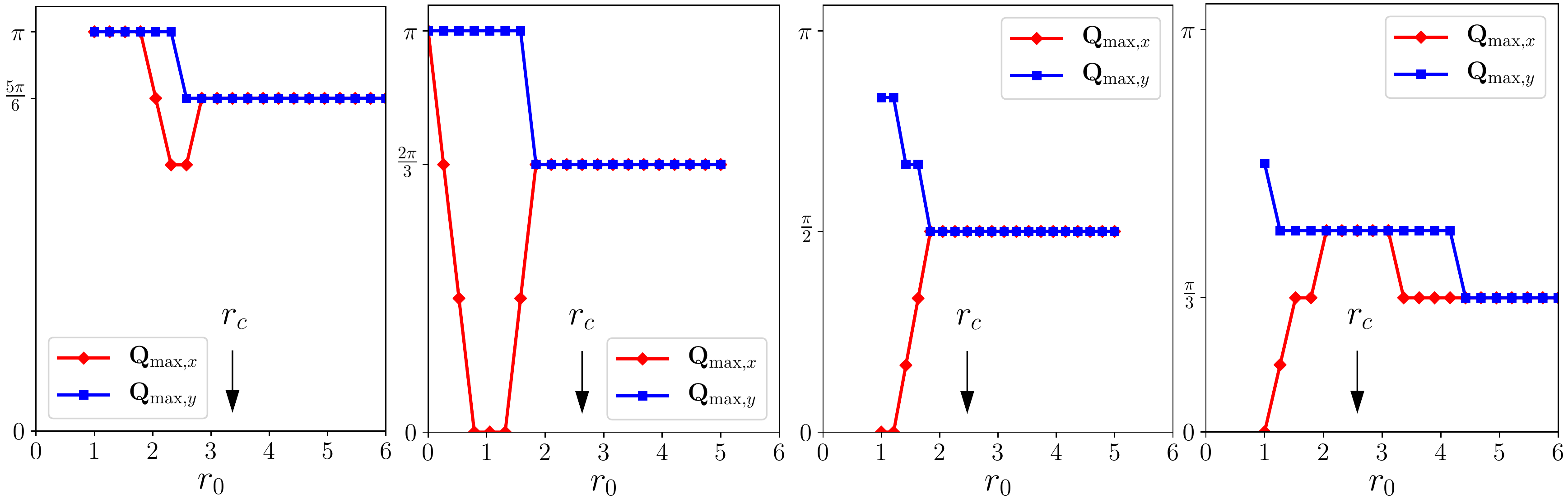}\caption{\label{fig:optCO}Maximal CO wave-vector $\mathbf{Q}_{\text{max}}=\left(Q_{x},Q_{y}\right)$
for one-dimensional band dispersions ($t_{y}=0$). From left to right:
$Q_{0}=5\pi/6$, $Q_{0}=2\pi/3$, $Q_{0}=\pi/2$, and $Q_{0}=\pi/3$.
The inverse temperature is set to $\beta t=14$. }
\end{figure}

For a given band dispersion, we have determined the wave-vector $\mathbf{Q}_{\text{max}}$
for which the CO susceptibility is maximal. In Fig. \ref{fig:optCO},
we present the coordinates of $\mathbf{Q}_{\text{max}}=\left(Q_{x},Q_{y}\right)$
as a function of $r_{0}$ for purely one-dimensional band dispersions
($t_{y}=0$) corresponding to $Q_{0}=5\pi/6$, $Q_{0}=2\pi/3$, $Q_{0}=\pi/2$,
and $Q_{0}=\pi/3$. The results are obtained for the inverse temperature
$\beta t=14$, and the qualitative features of the results are similar
when temperature is changed.

With the exception of $Q_{0}=\pi/3$, where the Fermi energy is small
compared to the bandwidth, the shift of the maximal CO wave-vector
to axial occurs inside the AFM phase, indicating that Fermi surface
reconstruction plays a major role. The maximal wave-vector deep inside
the AFM phase appears to be non-universal and dependent on the chemical
potential. For systems close to half-filling (i.e. $Q_{0}$ close
to $\pi$), the maximal wave-vector stays along the diagonal direction.
However, for systems farther away from half filling, the maximal wave-vector
shifts towards the axial direction. 

We attribute the qualitative differences among the maximal wave-vectors
of different chemical potentials to the location of the AFM hot spots.
For $Q_{0}=5\pi/6$ and $Q_{0}=2\pi/3$, the hot spots are located
near the diagonal directions. For $Q_{0}=\pi/2$ and $Q_{0}=\pi/3$,
however, the hot spots are closer to the antinodal region of the Fermi
surface. As a result, the AFM order can have a stronger effect on
the electronic states along the diagonal or axial directions depending
on the chemical potential. The diagonal-axial asymmetry of the reconstructed
Fermi surface is thus responsible for the position of the maximal
CO wave-vector. 

\subsection{Quasi-1D dispersions}

\begin{figure}
\includegraphics[width=1\linewidth]{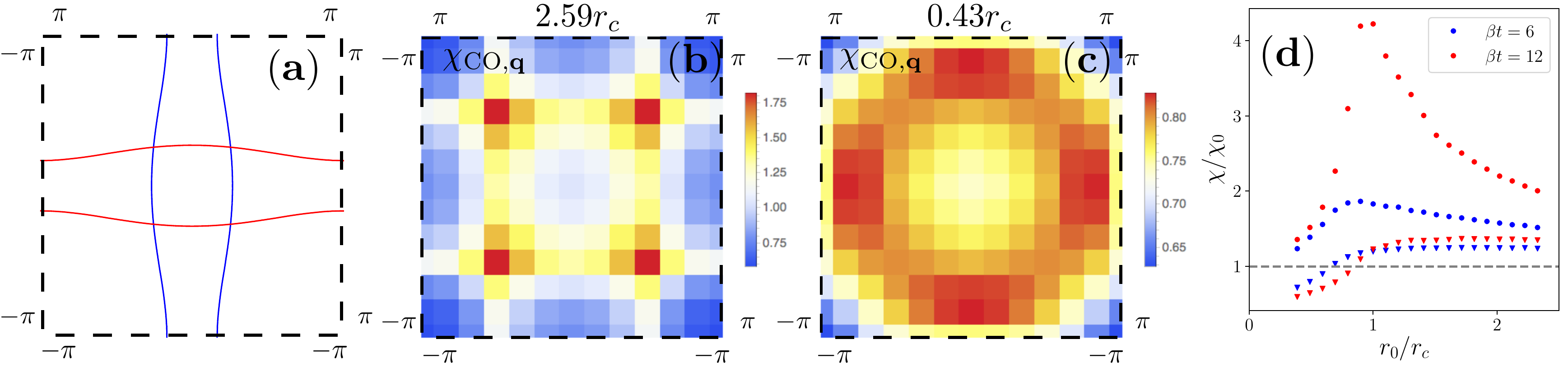}\caption{\label{fig:quasi1d}QMC results for the diagonal CO and SC susceptibilities
for the quasi-1D band dispersion plotted in (a). Panels (b) and (c)
show the CO susceptibility above and below the magnetic QCP, respectively.
Panel (d) shows the $r_{0}$ dependence of the SC (diamond) and diagonal-CO
(triangle) susceptibilties for temperatures $\beta t=6$ (blue) and
$\beta t=12$ (red), demonstrating the breaking of the SC-CO degeneracy
observed at half filling.}
\end{figure}

To show that the choice of one-dimensional band dispersions is not
responsible for the observed lifting of the SC-CO degeneracy and for
the observed shift of the CO wave-vector from diagonal to axial, here
we present results for a slightly curved band dispersion. In particular,
we consider the parameters $t_{y}=0.1t_{x}$ and $\mu=-\sqrt{3}+2t_{y}$.
This choice generates a small curvature to the Fermi surface while
retaining the momentum points $\left(\frac{\pi}{6},0\right)/\left(0,\frac{\pi}{6}\right)$
on the Fermi surface. In Fig. \ref{fig:quasi1d}, we present the band
dispersion for the non-interacting problem, the evolution of the CO
wave-vector as the AFM-QCP is crossed, as well as the dependence of
the sign-changing SC and diagonal CO susceptibilities as the AFM-QCP
is approached. The results are consistent with those presented in
the main text for 1D band dispersions.

\end{document}

%% file: cdw_final.bbl
%

%% file: cdw_final.bbl
\begin{thebibliography}{36}%
\makeatletter
\providecommand \@ifxundefined [1]{%
 \@ifx{#1\undefined}
}%
\providecommand \@ifnum [1]{%
 \ifnum #1\expandafter \@firstoftwo
 \else \expandafter \@secondoftwo
 \fi
}%
\providecommand \@ifx [1]{%
 \ifx #1\expandafter \@firstoftwo
 \else \expandafter \@secondoftwo
 \fi
}%
\providecommand \natexlab [1]{#1}%
\providecommand \enquote  [1]{``#1''}%
\providecommand \bibnamefont  [1]{#1}%
\providecommand \bibfnamefont [1]{#1}%
\providecommand \citenamefont [1]{#1}%
\providecommand \href@noop [0]{\@secondoftwo}%
\providecommand \href [0]{\begingroup \@sanitize@url \@href}%
\providecommand \@href[1]{\@@startlink{#1}\@@href}%
\providecommand \@@href[1]{\endgroup#1\@@endlink}%
\providecommand \@sanitize@url [0]{\catcode `\\12\catcode `\$12\catcode
  `\&12\catcode `\#12\catcode `\^12\catcode `\_12\catcode `\%12\relax}%
\providecommand \@@startlink[1]{}%
\providecommand \@@endlink[0]{}%
\providecommand \url  [0]{\begingroup\@sanitize@url \@url }%
\providecommand \@url [1]{\endgroup\@href {#1}{\urlprefix }}%
\providecommand \urlprefix  [0]{URL }%
\providecommand \Eprint [0]{\href }%
\providecommand \doibase [0]{http://dx.doi.org/}%
\providecommand \selectlanguage [0]{\@gobble}%
\providecommand \bibinfo  [0]{\@secondoftwo}%
\providecommand \bibfield  [0]{\@secondoftwo}%
\providecommand \translation [1]{[#1]}%
\providecommand \BibitemOpen [0]{}%
\providecommand \bibitemStop [0]{}%
\providecommand \bibitemNoStop [0]{.\EOS\space}%
\providecommand \EOS [0]{\spacefactor3000\relax}%
\providecommand \BibitemShut  [1]{\csname bibitem#1\endcsname}%
\let\auto@bib@innerbib\@empty
\bibitem [{\citenamefont {Scalapino}(2012)}]{Scalapino12}%
  \BibitemOpen
  \bibfield  {author} {\bibinfo {author} {\bibfnamefont {D.~J.}\ \bibnamefont
  {Scalapino}},\ }\href {\doibase 10.1103/RevModPhys.84.1383} {\bibfield
  {journal} {\bibinfo  {journal} {Rev. Mod. Phys.}\ }\textbf {\bibinfo {volume}
  {84}},\ \bibinfo {pages} {1383} (\bibinfo {year} {2012})}\BibitemShut
  {NoStop}%
\bibitem [{\citenamefont {Metlitski}\ and\ \citenamefont
  {Sachdev}(2010{\natexlab{a}})}]{Metlitski10b}%
  \BibitemOpen
  \bibfield  {author} {\bibinfo {author} {\bibfnamefont {M.~A.}\ \bibnamefont
  {Metlitski}}\ and\ \bibinfo {author} {\bibfnamefont {S.}~\bibnamefont
  {Sachdev}},\ }\href {http://stacks.iop.org/1367-2630/12/i=10/a=105007}
  {\bibfield  {journal} {\bibinfo  {journal} {New Journal of Physics}\ }\textbf
  {\bibinfo {volume} {12}},\ \bibinfo {pages} {105007} (\bibinfo {year}
  {2010}{\natexlab{a}})}\BibitemShut {NoStop}%
\bibitem [{\citenamefont {Metlitski}\ and\ \citenamefont
  {Sachdev}(2010{\natexlab{b}})}]{Metlitski10}%
  \BibitemOpen
  \bibfield  {author} {\bibinfo {author} {\bibfnamefont {M.~A.}\ \bibnamefont
  {Metlitski}}\ and\ \bibinfo {author} {\bibfnamefont {S.}~\bibnamefont
  {Sachdev}},\ }\href {\doibase 10.1103/PhysRevB.82.075128} {\bibfield
  {journal} {\bibinfo  {journal} {Phys. Rev. B}\ }\textbf {\bibinfo {volume}
  {82}},\ \bibinfo {pages} {075128} (\bibinfo {year}
  {2010}{\natexlab{b}})}\BibitemShut {NoStop}%
\bibitem [{\citenamefont {Efetov}\ \emph {et~al.}(2013)\citenamefont {Efetov},
  \citenamefont {Meier},\ and\ \citenamefont {Pepin}}]{Efetov13}%
  \BibitemOpen
  \bibfield  {author} {\bibinfo {author} {\bibfnamefont {K.~B.}\ \bibnamefont
  {Efetov}}, \bibinfo {author} {\bibfnamefont {H.}~\bibnamefont {Meier}}, \
  and\ \bibinfo {author} {\bibfnamefont {C.}~\bibnamefont {Pepin}},\ }\href
  {http://dx.doi.org/10.1038/nphys2641} {\bibfield  {journal} {\bibinfo
  {journal} {Nat Phys}\ }\textbf {\bibinfo {volume} {9}},\ \bibinfo {pages}
  {442} (\bibinfo {year} {2013})}\BibitemShut {NoStop}%
\bibitem [{\citenamefont {Wang}\ and\ \citenamefont
  {Chubukov}(2014)}]{Yuxuan14}%
  \BibitemOpen
  \bibfield  {author} {\bibinfo {author} {\bibfnamefont {Y.}~\bibnamefont
  {Wang}}\ and\ \bibinfo {author} {\bibfnamefont {A.}~\bibnamefont
  {Chubukov}},\ }\href {\doibase 10.1103/PhysRevB.90.035149} {\bibfield
  {journal} {\bibinfo  {journal} {Phys. Rev. B}\ }\textbf {\bibinfo {volume}
  {90}},\ \bibinfo {pages} {035149} (\bibinfo {year} {2014})}\BibitemShut
  {NoStop}%
\bibitem [{\citenamefont {Allais}\ \emph {et~al.}(2014)\citenamefont {Allais},
  \citenamefont {Bauer},\ and\ \citenamefont {Sachdev}}]{Allais14}%
  \BibitemOpen
  \bibfield  {author} {\bibinfo {author} {\bibfnamefont {A.}~\bibnamefont
  {Allais}}, \bibinfo {author} {\bibfnamefont {J.}~\bibnamefont {Bauer}}, \
  and\ \bibinfo {author} {\bibfnamefont {S.}~\bibnamefont {Sachdev}},\ }\href
  {\doibase 10.1103/PhysRevB.90.155114} {\bibfield  {journal} {\bibinfo
  {journal} {Phys. Rev. B}\ }\textbf {\bibinfo {volume} {90}},\ \bibinfo
  {pages} {155114} (\bibinfo {year} {2014})}\BibitemShut {NoStop}%
\bibitem [{\citenamefont {Wu}\ \emph {et~al.}(2011)\citenamefont {Wu},
  \citenamefont {Mayaffre}, \citenamefont {Kramer}, \citenamefont {Horvatic},
  \citenamefont {Berthier}, \citenamefont {Hardy}, \citenamefont {Liang},
  \citenamefont {Bonn},\ and\ \citenamefont {Julien}}]{Wu11}%
  \BibitemOpen
  \bibfield  {author} {\bibinfo {author} {\bibfnamefont {T.}~\bibnamefont
  {Wu}}, \bibinfo {author} {\bibfnamefont {H.}~\bibnamefont {Mayaffre}},
  \bibinfo {author} {\bibfnamefont {S.}~\bibnamefont {Kramer}}, \bibinfo
  {author} {\bibfnamefont {M.}~\bibnamefont {Horvatic}}, \bibinfo {author}
  {\bibfnamefont {C.}~\bibnamefont {Berthier}}, \bibinfo {author}
  {\bibfnamefont {W.~N.}\ \bibnamefont {Hardy}}, \bibinfo {author}
  {\bibfnamefont {R.}~\bibnamefont {Liang}}, \bibinfo {author} {\bibfnamefont
  {D.~A.}\ \bibnamefont {Bonn}}, \ and\ \bibinfo {author} {\bibfnamefont
  {M.-H.}\ \bibnamefont {Julien}},\ }\href
  {http://dx.doi.org/10.1038/nature10345} {\bibfield  {journal} {\bibinfo
  {journal} {Nature}\ }\textbf {\bibinfo {volume} {477}},\ \bibinfo {pages}
  {191} (\bibinfo {year} {2011})}\BibitemShut {NoStop}%
\bibitem [{\citenamefont {Chang}\ \emph {et~al.}(2012)\citenamefont {Chang},
  \citenamefont {Blackburn}, \citenamefont {Holmes}, \citenamefont
  {Christensen}, \citenamefont {Larsen}, \citenamefont {Mesot}, \citenamefont
  {Liang}, \citenamefont {Bonn}, \citenamefont {Hardy}, \citenamefont
  {Watenphul}, \citenamefont {Zimmermann}, \citenamefont {Forgan},\ and\
  \citenamefont {Hayden}}]{Chang12}%
  \BibitemOpen
  \bibfield  {author} {\bibinfo {author} {\bibfnamefont {J.}~\bibnamefont
  {Chang}}, \bibinfo {author} {\bibfnamefont {E.}~\bibnamefont {Blackburn}},
  \bibinfo {author} {\bibfnamefont {A.~T.}\ \bibnamefont {Holmes}}, \bibinfo
  {author} {\bibfnamefont {N.~B.}\ \bibnamefont {Christensen}}, \bibinfo
  {author} {\bibfnamefont {J.}~\bibnamefont {Larsen}}, \bibinfo {author}
  {\bibfnamefont {J.}~\bibnamefont {Mesot}}, \bibinfo {author} {\bibfnamefont
  {R.}~\bibnamefont {Liang}}, \bibinfo {author} {\bibfnamefont {D.~A.}\
  \bibnamefont {Bonn}}, \bibinfo {author} {\bibfnamefont {W.~N.}\ \bibnamefont
  {Hardy}}, \bibinfo {author} {\bibfnamefont {A.}~\bibnamefont {Watenphul}},
  \bibinfo {author} {\bibfnamefont {M.~v.}\ \bibnamefont {Zimmermann}},
  \bibinfo {author} {\bibfnamefont {E.~M.}\ \bibnamefont {Forgan}}, \ and\
  \bibinfo {author} {\bibfnamefont {S.~M.}\ \bibnamefont {Hayden}},\ }\href
  {http://dx.doi.org/10.1038/nphys2456} {\bibfield  {journal} {\bibinfo
  {journal} {Nat Phys}\ }\textbf {\bibinfo {volume} {8}},\ \bibinfo {pages}
  {871} (\bibinfo {year} {2012})}\BibitemShut {NoStop}%
\bibitem [{\citenamefont {Achkar}\ \emph {et~al.}(2012)\citenamefont {Achkar},
  \citenamefont {Sutarto}, \citenamefont {Mao}, \citenamefont {He},
  \citenamefont {Frano}, \citenamefont {Blanco-Canosa}, \citenamefont
  {Le~Tacon}, \citenamefont {Ghiringhelli}, \citenamefont {Braicovich},
  \citenamefont {Minola}, \citenamefont {Moretti~Sala}, \citenamefont
  {Mazzoli}, \citenamefont {Liang}, \citenamefont {Bonn}, \citenamefont
  {Hardy}, \citenamefont {Keimer}, \citenamefont {Sawatzky},\ and\
  \citenamefont {Hawthorn}}]{Achkar12}%
  \BibitemOpen
  \bibfield  {author} {\bibinfo {author} {\bibfnamefont {A.~J.}\ \bibnamefont
  {Achkar}}, \bibinfo {author} {\bibfnamefont {R.}~\bibnamefont {Sutarto}},
  \bibinfo {author} {\bibfnamefont {X.}~\bibnamefont {Mao}}, \bibinfo {author}
  {\bibfnamefont {F.}~\bibnamefont {He}}, \bibinfo {author} {\bibfnamefont
  {A.}~\bibnamefont {Frano}}, \bibinfo {author} {\bibfnamefont
  {S.}~\bibnamefont {Blanco-Canosa}}, \bibinfo {author} {\bibfnamefont
  {M.}~\bibnamefont {Le~Tacon}}, \bibinfo {author} {\bibfnamefont
  {G.}~\bibnamefont {Ghiringhelli}}, \bibinfo {author} {\bibfnamefont
  {L.}~\bibnamefont {Braicovich}}, \bibinfo {author} {\bibfnamefont
  {M.}~\bibnamefont {Minola}}, \bibinfo {author} {\bibfnamefont
  {M.}~\bibnamefont {Moretti~Sala}}, \bibinfo {author} {\bibfnamefont
  {C.}~\bibnamefont {Mazzoli}}, \bibinfo {author} {\bibfnamefont
  {R.}~\bibnamefont {Liang}}, \bibinfo {author} {\bibfnamefont {D.~A.}\
  \bibnamefont {Bonn}}, \bibinfo {author} {\bibfnamefont {W.~N.}\ \bibnamefont
  {Hardy}}, \bibinfo {author} {\bibfnamefont {B.}~\bibnamefont {Keimer}},
  \bibinfo {author} {\bibfnamefont {G.~A.}\ \bibnamefont {Sawatzky}}, \ and\
  \bibinfo {author} {\bibfnamefont {D.~G.}\ \bibnamefont {Hawthorn}},\ }\href
  {\doibase 10.1103/PhysRevLett.109.167001} {\bibfield  {journal} {\bibinfo
  {journal} {Phys. Rev. Lett.}\ }\textbf {\bibinfo {volume} {109}},\ \bibinfo
  {pages} {167001} (\bibinfo {year} {2012})}\BibitemShut {NoStop}%
\bibitem [{\citenamefont {Ghiringhelli}\ \emph {et~al.}(2012)\citenamefont
  {Ghiringhelli}, \citenamefont {Le~Tacon}, \citenamefont {Minola},
  \citenamefont {Blanco-Canosa}, \citenamefont {Mazzoli}, \citenamefont
  {Brookes}, \citenamefont {De~Luca}, \citenamefont {Frano}, \citenamefont
  {Hawthorn}, \citenamefont {He}, \citenamefont {Loew}, \citenamefont {Sala},
  \citenamefont {Peets}, \citenamefont {Salluzzo}, \citenamefont {Schierle},
  \citenamefont {Sutarto}, \citenamefont {Sawatzky}, \citenamefont {Weschke},
  \citenamefont {Keimer},\ and\ \citenamefont {Braicovich}}]{Ghiringhelli12}%
  \BibitemOpen
  \bibfield  {author} {\bibinfo {author} {\bibfnamefont {G.}~\bibnamefont
  {Ghiringhelli}}, \bibinfo {author} {\bibfnamefont {M.}~\bibnamefont
  {Le~Tacon}}, \bibinfo {author} {\bibfnamefont {M.}~\bibnamefont {Minola}},
  \bibinfo {author} {\bibfnamefont {S.}~\bibnamefont {Blanco-Canosa}}, \bibinfo
  {author} {\bibfnamefont {C.}~\bibnamefont {Mazzoli}}, \bibinfo {author}
  {\bibfnamefont {N.~B.}\ \bibnamefont {Brookes}}, \bibinfo {author}
  {\bibfnamefont {G.~M.}\ \bibnamefont {De~Luca}}, \bibinfo {author}
  {\bibfnamefont {A.}~\bibnamefont {Frano}}, \bibinfo {author} {\bibfnamefont
  {D.~G.}\ \bibnamefont {Hawthorn}}, \bibinfo {author} {\bibfnamefont
  {F.}~\bibnamefont {He}}, \bibinfo {author} {\bibfnamefont {T.}~\bibnamefont
  {Loew}}, \bibinfo {author} {\bibfnamefont {M.~M.}\ \bibnamefont {Sala}},
  \bibinfo {author} {\bibfnamefont {D.~C.}\ \bibnamefont {Peets}}, \bibinfo
  {author} {\bibfnamefont {M.}~\bibnamefont {Salluzzo}}, \bibinfo {author}
  {\bibfnamefont {E.}~\bibnamefont {Schierle}}, \bibinfo {author}
  {\bibfnamefont {R.}~\bibnamefont {Sutarto}}, \bibinfo {author} {\bibfnamefont
  {G.~A.}\ \bibnamefont {Sawatzky}}, \bibinfo {author} {\bibfnamefont
  {E.}~\bibnamefont {Weschke}}, \bibinfo {author} {\bibfnamefont
  {B.}~\bibnamefont {Keimer}}, \ and\ \bibinfo {author} {\bibfnamefont
  {L.}~\bibnamefont {Braicovich}},\ }\href {\doibase 10.1126/science.1223532}
  {\bibfield  {journal} {\bibinfo  {journal} {Science}\ }\textbf {\bibinfo
  {volume} {337}},\ \bibinfo {pages} {821} (\bibinfo {year}
  {2012})}\BibitemShut {NoStop}%
\bibitem [{\citenamefont {Blackburn}\ \emph {et~al.}(2013)\citenamefont
  {Blackburn}, \citenamefont {Chang}, \citenamefont {H\"ucker}, \citenamefont
  {Holmes}, \citenamefont {Christensen}, \citenamefont {Liang}, \citenamefont
  {Bonn}, \citenamefont {Hardy}, \citenamefont {R\"utt}, \citenamefont
  {Gutowski}, \citenamefont {Zimmermann}, \citenamefont {Forgan},\ and\
  \citenamefont {Hayden}}]{Blackburn13}%
  \BibitemOpen
  \bibfield  {author} {\bibinfo {author} {\bibfnamefont {E.}~\bibnamefont
  {Blackburn}}, \bibinfo {author} {\bibfnamefont {J.}~\bibnamefont {Chang}},
  \bibinfo {author} {\bibfnamefont {M.}~\bibnamefont {H\"ucker}}, \bibinfo
  {author} {\bibfnamefont {A.~T.}\ \bibnamefont {Holmes}}, \bibinfo {author}
  {\bibfnamefont {N.~B.}\ \bibnamefont {Christensen}}, \bibinfo {author}
  {\bibfnamefont {R.}~\bibnamefont {Liang}}, \bibinfo {author} {\bibfnamefont
  {D.~A.}\ \bibnamefont {Bonn}}, \bibinfo {author} {\bibfnamefont {W.~N.}\
  \bibnamefont {Hardy}}, \bibinfo {author} {\bibfnamefont {U.}~\bibnamefont
  {R\"utt}}, \bibinfo {author} {\bibfnamefont {O.}~\bibnamefont {Gutowski}},
  \bibinfo {author} {\bibfnamefont {M.~v.}\ \bibnamefont {Zimmermann}},
  \bibinfo {author} {\bibfnamefont {E.~M.}\ \bibnamefont {Forgan}}, \ and\
  \bibinfo {author} {\bibfnamefont {S.~M.}\ \bibnamefont {Hayden}},\ }\href
  {\doibase 10.1103/PhysRevLett.110.137004} {\bibfield  {journal} {\bibinfo
  {journal} {Phys. Rev. Lett.}\ }\textbf {\bibinfo {volume} {110}},\ \bibinfo
  {pages} {137004} (\bibinfo {year} {2013})}\BibitemShut {NoStop}%
\bibitem [{\citenamefont {Doiron-Leyraud}\ \emph {et~al.}(2013)\citenamefont
  {Doiron-Leyraud}, \citenamefont {Lepault}, \citenamefont {Cyr-Choini\`ere},
  \citenamefont {Vignolle}, \citenamefont {Grissonnanche}, \citenamefont
  {Lalibert\'e}, \citenamefont {Chang}, \citenamefont {Bari\ifmmode
  \check{s}\else \v{s}\fi{}i\ifmmode~\acute{c}\else \'{c}\fi{}}, \citenamefont
  {Chan}, \citenamefont {Ji}, \citenamefont {Zhao}, \citenamefont {Li},
  \citenamefont {Greven}, \citenamefont {Proust},\ and\ \citenamefont
  {Taillefer}}]{Leyraud13}%
  \BibitemOpen
  \bibfield  {author} {\bibinfo {author} {\bibfnamefont {N.}~\bibnamefont
  {Doiron-Leyraud}}, \bibinfo {author} {\bibfnamefont {S.}~\bibnamefont
  {Lepault}}, \bibinfo {author} {\bibfnamefont {O.}~\bibnamefont
  {Cyr-Choini\`ere}}, \bibinfo {author} {\bibfnamefont {B.}~\bibnamefont
  {Vignolle}}, \bibinfo {author} {\bibfnamefont {G.}~\bibnamefont
  {Grissonnanche}}, \bibinfo {author} {\bibfnamefont {F.}~\bibnamefont
  {Lalibert\'e}}, \bibinfo {author} {\bibfnamefont {J.}~\bibnamefont {Chang}},
  \bibinfo {author} {\bibfnamefont {N.}~\bibnamefont {Bari\ifmmode
  \check{s}\else \v{s}\fi{}i\ifmmode~\acute{c}\else \'{c}\fi{}}}, \bibinfo
  {author} {\bibfnamefont {M.~K.}\ \bibnamefont {Chan}}, \bibinfo {author}
  {\bibfnamefont {L.}~\bibnamefont {Ji}}, \bibinfo {author} {\bibfnamefont
  {X.}~\bibnamefont {Zhao}}, \bibinfo {author} {\bibfnamefont {Y.}~\bibnamefont
  {Li}}, \bibinfo {author} {\bibfnamefont {M.}~\bibnamefont {Greven}}, \bibinfo
  {author} {\bibfnamefont {C.}~\bibnamefont {Proust}}, \ and\ \bibinfo {author}
  {\bibfnamefont {L.}~\bibnamefont {Taillefer}},\ }\href {\doibase
  10.1103/PhysRevX.3.021019} {\bibfield  {journal} {\bibinfo  {journal} {Phys.
  Rev. X}\ }\textbf {\bibinfo {volume} {3}},\ \bibinfo {pages} {021019}
  (\bibinfo {year} {2013})}\BibitemShut {NoStop}%
\bibitem [{\citenamefont {LeBoeuf}\ \emph {et~al.}(2013)\citenamefont
  {LeBoeuf}, \citenamefont {Kramer}, \citenamefont {Hardy}, \citenamefont
  {Liang}, \citenamefont {Bonn},\ and\ \citenamefont {Proust}}]{LeBoeuf13}%
  \BibitemOpen
  \bibfield  {author} {\bibinfo {author} {\bibfnamefont {D.}~\bibnamefont
  {LeBoeuf}}, \bibinfo {author} {\bibfnamefont {S.}~\bibnamefont {Kramer}},
  \bibinfo {author} {\bibfnamefont {W.~N.}\ \bibnamefont {Hardy}}, \bibinfo
  {author} {\bibfnamefont {R.}~\bibnamefont {Liang}}, \bibinfo {author}
  {\bibfnamefont {D.~A.}\ \bibnamefont {Bonn}}, \ and\ \bibinfo {author}
  {\bibfnamefont {C.}~\bibnamefont {Proust}},\ }\href
  {http://dx.doi.org/10.1038/nphys2502} {\bibfield  {journal} {\bibinfo
  {journal} {Nat Phys}\ }\textbf {\bibinfo {volume} {9}},\ \bibinfo {pages}
  {79} (\bibinfo {year} {2013})}\BibitemShut {NoStop}%
\bibitem [{\citenamefont {Comin}\ \emph {et~al.}(2014)\citenamefont {Comin},
  \citenamefont {Frano}, \citenamefont {Yee}, \citenamefont {Yoshida},
  \citenamefont {Eisaki}, \citenamefont {Schierle}, \citenamefont {Weschke},
  \citenamefont {Sutarto}, \citenamefont {He}, \citenamefont {Soumyanarayanan},
  \citenamefont {He}, \citenamefont {Le~Tacon}, \citenamefont {Elfimov},
  \citenamefont {Hoffman}, \citenamefont {Sawatzky}, \citenamefont {Keimer},\
  and\ \citenamefont {Damascelli}}]{Comin14}%
  \BibitemOpen
  \bibfield  {author} {\bibinfo {author} {\bibfnamefont {R.}~\bibnamefont
  {Comin}}, \bibinfo {author} {\bibfnamefont {A.}~\bibnamefont {Frano}},
  \bibinfo {author} {\bibfnamefont {M.~M.}\ \bibnamefont {Yee}}, \bibinfo
  {author} {\bibfnamefont {Y.}~\bibnamefont {Yoshida}}, \bibinfo {author}
  {\bibfnamefont {H.}~\bibnamefont {Eisaki}}, \bibinfo {author} {\bibfnamefont
  {E.}~\bibnamefont {Schierle}}, \bibinfo {author} {\bibfnamefont
  {E.}~\bibnamefont {Weschke}}, \bibinfo {author} {\bibfnamefont
  {R.}~\bibnamefont {Sutarto}}, \bibinfo {author} {\bibfnamefont
  {F.}~\bibnamefont {He}}, \bibinfo {author} {\bibfnamefont {A.}~\bibnamefont
  {Soumyanarayanan}}, \bibinfo {author} {\bibfnamefont {Y.}~\bibnamefont {He}},
  \bibinfo {author} {\bibfnamefont {M.}~\bibnamefont {Le~Tacon}}, \bibinfo
  {author} {\bibfnamefont {I.~S.}\ \bibnamefont {Elfimov}}, \bibinfo {author}
  {\bibfnamefont {J.~E.}\ \bibnamefont {Hoffman}}, \bibinfo {author}
  {\bibfnamefont {G.~A.}\ \bibnamefont {Sawatzky}}, \bibinfo {author}
  {\bibfnamefont {B.}~\bibnamefont {Keimer}}, \ and\ \bibinfo {author}
  {\bibfnamefont {A.}~\bibnamefont {Damascelli}},\ }\href {\doibase
  10.1126/science.1242996} {\bibfield  {journal} {\bibinfo  {journal}
  {Science}\ }\textbf {\bibinfo {volume} {343}},\ \bibinfo {pages} {390}
  (\bibinfo {year} {2014})}\BibitemShut {NoStop}%
\bibitem [{\citenamefont {Fujita}\ \emph {et~al.}(2014)\citenamefont {Fujita},
  \citenamefont {Kim}, \citenamefont {Lee}, \citenamefont {Lee}, \citenamefont
  {Hamidian}, \citenamefont {Firmo}, \citenamefont {Mukhopadhyay},
  \citenamefont {Eisaki}, \citenamefont {Uchida}, \citenamefont {Lawler},
  \citenamefont {Kim},\ and\ \citenamefont {Davis}}]{Fujita14}%
  \BibitemOpen
  \bibfield  {author} {\bibinfo {author} {\bibfnamefont {K.}~\bibnamefont
  {Fujita}}, \bibinfo {author} {\bibfnamefont {C.~K.}\ \bibnamefont {Kim}},
  \bibinfo {author} {\bibfnamefont {I.}~\bibnamefont {Lee}}, \bibinfo {author}
  {\bibfnamefont {J.}~\bibnamefont {Lee}}, \bibinfo {author} {\bibfnamefont
  {M.~H.}\ \bibnamefont {Hamidian}}, \bibinfo {author} {\bibfnamefont {I.~A.}\
  \bibnamefont {Firmo}}, \bibinfo {author} {\bibfnamefont {S.}~\bibnamefont
  {Mukhopadhyay}}, \bibinfo {author} {\bibfnamefont {H.}~\bibnamefont
  {Eisaki}}, \bibinfo {author} {\bibfnamefont {S.}~\bibnamefont {Uchida}},
  \bibinfo {author} {\bibfnamefont {M.~J.}\ \bibnamefont {Lawler}}, \bibinfo
  {author} {\bibfnamefont {E.-A.}\ \bibnamefont {Kim}}, \ and\ \bibinfo
  {author} {\bibfnamefont {J.~C.}\ \bibnamefont {Davis}},\ }\href {\doibase
  10.1126/science.1248783} {\bibfield  {journal} {\bibinfo  {journal}
  {Science}\ }\textbf {\bibinfo {volume} {344}},\ \bibinfo {pages} {612}
  (\bibinfo {year} {2014})}\BibitemShut {NoStop}%
\bibitem [{\citenamefont {da~Silva~Neto}\ \emph {et~al.}(2014)\citenamefont
  {da~Silva~Neto}, \citenamefont {Aynajian}, \citenamefont {Frano},
  \citenamefont {Comin}, \citenamefont {Schierle}, \citenamefont {Weschke},
  \citenamefont {Gyenis}, \citenamefont {Wen}, \citenamefont {Schneeloch},
  \citenamefont {Xu}, \citenamefont {Ono}, \citenamefont {Gu}, \citenamefont
  {Le~Tacon},\ and\ \citenamefont {Yazdani}}]{Neto14}%
  \BibitemOpen
  \bibfield  {author} {\bibinfo {author} {\bibfnamefont {E.~H.}\ \bibnamefont
  {da~Silva~Neto}}, \bibinfo {author} {\bibfnamefont {P.}~\bibnamefont
  {Aynajian}}, \bibinfo {author} {\bibfnamefont {A.}~\bibnamefont {Frano}},
  \bibinfo {author} {\bibfnamefont {R.}~\bibnamefont {Comin}}, \bibinfo
  {author} {\bibfnamefont {E.}~\bibnamefont {Schierle}}, \bibinfo {author}
  {\bibfnamefont {E.}~\bibnamefont {Weschke}}, \bibinfo {author} {\bibfnamefont
  {A.}~\bibnamefont {Gyenis}}, \bibinfo {author} {\bibfnamefont
  {J.}~\bibnamefont {Wen}}, \bibinfo {author} {\bibfnamefont {J.}~\bibnamefont
  {Schneeloch}}, \bibinfo {author} {\bibfnamefont {Z.}~\bibnamefont {Xu}},
  \bibinfo {author} {\bibfnamefont {S.}~\bibnamefont {Ono}}, \bibinfo {author}
  {\bibfnamefont {G.}~\bibnamefont {Gu}}, \bibinfo {author} {\bibfnamefont
  {M.}~\bibnamefont {Le~Tacon}}, \ and\ \bibinfo {author} {\bibfnamefont
  {A.}~\bibnamefont {Yazdani}},\ }\href
  {http://science.sciencemag.org/content/343/6169/393.abstract} {\bibfield
  {journal} {\bibinfo  {journal} {Science}\ }\textbf {\bibinfo {volume}
  {343}},\ \bibinfo {pages} {393} (\bibinfo {year} {2014})}\BibitemShut
  {NoStop}%
\bibitem [{\citenamefont {Mesaros}\ \emph {et~al.}(2016)\citenamefont
  {Mesaros}, \citenamefont {Fujita}, \citenamefont {Edkins}, \citenamefont
  {Hamidian}, \citenamefont {Eisaki}, \citenamefont {Uchida}, \citenamefont
  {Davis}, \citenamefont {Lawler},\ and\ \citenamefont {Kim}}]{Mesaros16}%
  \BibitemOpen
  \bibfield  {author} {\bibinfo {author} {\bibfnamefont {A.}~\bibnamefont
  {Mesaros}}, \bibinfo {author} {\bibfnamefont {K.}~\bibnamefont {Fujita}},
  \bibinfo {author} {\bibfnamefont {S.~D.}\ \bibnamefont {Edkins}}, \bibinfo
  {author} {\bibfnamefont {M.~H.}\ \bibnamefont {Hamidian}}, \bibinfo {author}
  {\bibfnamefont {H.}~\bibnamefont {Eisaki}}, \bibinfo {author} {\bibfnamefont
  {S.-i.}\ \bibnamefont {Uchida}}, \bibinfo {author} {\bibfnamefont {J.~C.~S.}\
  \bibnamefont {Davis}}, \bibinfo {author} {\bibfnamefont {M.~J.}\ \bibnamefont
  {Lawler}}, \ and\ \bibinfo {author} {\bibfnamefont {E.-A.}\ \bibnamefont
  {Kim}},\ }\href {\doibase 10.1073/pnas.1614247113} {\bibfield  {journal}
  {\bibinfo  {journal} {Proceedings of the National Academy of Sciences}\
  }\textbf {\bibinfo {volume} {113}},\ \bibinfo {pages} {12661} (\bibinfo
  {year} {2016})}\BibitemShut {NoStop}%
\bibitem [{\citenamefont {Jang}\ \emph {et~al.}(2016)\citenamefont {Jang},
  \citenamefont {Lee}, \citenamefont {Nojiri}, \citenamefont {Matsuzawa},
  \citenamefont {Yasumura}, \citenamefont {Nie}, \citenamefont {Maharaj},
  \citenamefont {Gerber}, \citenamefont {Liu}, \citenamefont {Mehta},
  \citenamefont {Bonn}, \citenamefont {Liang}, \citenamefont {Hardy},
  \citenamefont {Burns}, \citenamefont {Islam}, \citenamefont {Song},
  \citenamefont {Hastings}, \citenamefont {Devereaux}, \citenamefont {Shen},
  \citenamefont {Kivelson}, \citenamefont {Kao}, \citenamefont {Zhu},\ and\
  \citenamefont {Lee}}]{Jang16}%
  \BibitemOpen
  \bibfield  {author} {\bibinfo {author} {\bibfnamefont {H.}~\bibnamefont
  {Jang}}, \bibinfo {author} {\bibfnamefont {W.-S.}\ \bibnamefont {Lee}},
  \bibinfo {author} {\bibfnamefont {H.}~\bibnamefont {Nojiri}}, \bibinfo
  {author} {\bibfnamefont {S.}~\bibnamefont {Matsuzawa}}, \bibinfo {author}
  {\bibfnamefont {H.}~\bibnamefont {Yasumura}}, \bibinfo {author}
  {\bibfnamefont {L.}~\bibnamefont {Nie}}, \bibinfo {author} {\bibfnamefont
  {A.~V.}\ \bibnamefont {Maharaj}}, \bibinfo {author} {\bibfnamefont
  {S.}~\bibnamefont {Gerber}}, \bibinfo {author} {\bibfnamefont {Y.-J.}\
  \bibnamefont {Liu}}, \bibinfo {author} {\bibfnamefont {A.}~\bibnamefont
  {Mehta}}, \bibinfo {author} {\bibfnamefont {D.~A.}\ \bibnamefont {Bonn}},
  \bibinfo {author} {\bibfnamefont {R.}~\bibnamefont {Liang}}, \bibinfo
  {author} {\bibfnamefont {W.~N.}\ \bibnamefont {Hardy}}, \bibinfo {author}
  {\bibfnamefont {C.~A.}\ \bibnamefont {Burns}}, \bibinfo {author}
  {\bibfnamefont {Z.}~\bibnamefont {Islam}}, \bibinfo {author} {\bibfnamefont
  {S.}~\bibnamefont {Song}}, \bibinfo {author} {\bibfnamefont {J.}~\bibnamefont
  {Hastings}}, \bibinfo {author} {\bibfnamefont {T.~P.}\ \bibnamefont
  {Devereaux}}, \bibinfo {author} {\bibfnamefont {Z.-X.}\ \bibnamefont {Shen}},
  \bibinfo {author} {\bibfnamefont {S.~A.}\ \bibnamefont {Kivelson}}, \bibinfo
  {author} {\bibfnamefont {C.-C.}\ \bibnamefont {Kao}}, \bibinfo {author}
  {\bibfnamefont {D.}~\bibnamefont {Zhu}}, \ and\ \bibinfo {author}
  {\bibfnamefont {J.-S.}\ \bibnamefont {Lee}},\ }\href {\doibase
  10.1073/pnas.1612849113} {\bibfield  {journal} {\bibinfo  {journal}
  {Proceedings of the National Academy of Sciences}\ }\textbf {\bibinfo
  {volume} {113}},\ \bibinfo {pages} {14645} (\bibinfo {year}
  {2016})}\BibitemShut {NoStop}%
\bibitem [{\citenamefont {Chang}\ \emph {et~al.}(2016)\citenamefont {Chang},
  \citenamefont {Blackburn}, \citenamefont {Ivashko}, \citenamefont {Holmes},
  \citenamefont {Christensen}, \citenamefont {H\"ucker}, \citenamefont {Liang},
  \citenamefont {Bonn}, \citenamefont {Hardy}, \citenamefont {R\"utt},
  \citenamefont {Zimmermann}, \citenamefont {Forgan},\ and\ \citenamefont
  {Hayden}}]{Chang16}%
  \BibitemOpen
  \bibfield  {author} {\bibinfo {author} {\bibfnamefont {J.}~\bibnamefont
  {Chang}}, \bibinfo {author} {\bibfnamefont {E.}~\bibnamefont {Blackburn}},
  \bibinfo {author} {\bibfnamefont {O.}~\bibnamefont {Ivashko}}, \bibinfo
  {author} {\bibfnamefont {A.~T.}\ \bibnamefont {Holmes}}, \bibinfo {author}
  {\bibfnamefont {N.~B.}\ \bibnamefont {Christensen}}, \bibinfo {author}
  {\bibfnamefont {M.}~\bibnamefont {H\"ucker}}, \bibinfo {author}
  {\bibfnamefont {R.}~\bibnamefont {Liang}}, \bibinfo {author} {\bibfnamefont
  {D.~A.}\ \bibnamefont {Bonn}}, \bibinfo {author} {\bibfnamefont {W.~N.}\
  \bibnamefont {Hardy}}, \bibinfo {author} {\bibfnamefont {U.}~\bibnamefont
  {R\"utt}}, \bibinfo {author} {\bibfnamefont {M.~v.}\ \bibnamefont
  {Zimmermann}}, \bibinfo {author} {\bibfnamefont {E.~M.}\ \bibnamefont
  {Forgan}}, \ and\ \bibinfo {author} {\bibfnamefont {S.~M.}\ \bibnamefont
  {Hayden}},\ }\href {http://dx.doi.org/10.1038/ncomms11494} {\bibfield
  {journal} {\bibinfo  {journal} {Nature Communications}\ }\textbf {\bibinfo
  {volume} {7}},\ \bibinfo {pages} {11494} (\bibinfo {year}
  {2016})}\BibitemShut {NoStop}%
\bibitem [{\citenamefont {Bulut}\ \emph {et~al.}(2013)\citenamefont {Bulut},
  \citenamefont {Atkinson},\ and\ \citenamefont {Kampf}}]{Bulut13}%
  \BibitemOpen
  \bibfield  {author} {\bibinfo {author} {\bibfnamefont {S.}~\bibnamefont
  {Bulut}}, \bibinfo {author} {\bibfnamefont {W.~A.}\ \bibnamefont {Atkinson}},
  \ and\ \bibinfo {author} {\bibfnamefont {A.~P.}\ \bibnamefont {Kampf}},\
  }\href {\doibase 10.1103/PhysRevB.88.155132} {\bibfield  {journal} {\bibinfo
  {journal} {Phys. Rev. B}\ }\textbf {\bibinfo {volume} {88}},\ \bibinfo
  {pages} {155132} (\bibinfo {year} {2013})}\BibitemShut {NoStop}%
\bibitem [{\citenamefont {Sachdev}\ and\ \citenamefont
  {La~Placa}(2013)}]{Sachdev13}%
  \BibitemOpen
  \bibfield  {author} {\bibinfo {author} {\bibfnamefont {S.}~\bibnamefont
  {Sachdev}}\ and\ \bibinfo {author} {\bibfnamefont {R.}~\bibnamefont
  {La~Placa}},\ }\href {\doibase 10.1103/PhysRevLett.111.027202} {\bibfield
  {journal} {\bibinfo  {journal} {Phys. Rev. Lett.}\ }\textbf {\bibinfo
  {volume} {111}},\ \bibinfo {pages} {027202} (\bibinfo {year}
  {2013})}\BibitemShut {NoStop}%
\bibitem [{\citenamefont {Hayward}\ \emph {et~al.}(2014)\citenamefont
  {Hayward}, \citenamefont {Hawthorn}, \citenamefont {Melko},\ and\
  \citenamefont {Sachdev}}]{Hayward14}%
  \BibitemOpen
  \bibfield  {author} {\bibinfo {author} {\bibfnamefont {L.~E.}\ \bibnamefont
  {Hayward}}, \bibinfo {author} {\bibfnamefont {D.~G.}\ \bibnamefont
  {Hawthorn}}, \bibinfo {author} {\bibfnamefont {R.~G.}\ \bibnamefont {Melko}},
  \ and\ \bibinfo {author} {\bibfnamefont {S.}~\bibnamefont {Sachdev}},\ }\href
  {\doibase 10.1126/science.1246310} {\bibfield  {journal} {\bibinfo  {journal}
  {Science}\ }\textbf {\bibinfo {volume} {343}},\ \bibinfo {pages} {1336}
  (\bibinfo {year} {2014})}\BibitemShut {NoStop}%
\bibitem [{\citenamefont {Caprara}\ \emph {et~al.}(2017)\citenamefont
  {Caprara}, \citenamefont {Di~Castro}, \citenamefont {Seibold},\ and\
  \citenamefont {Grilli}}]{Grilli17}%
  \BibitemOpen
  \bibfield  {author} {\bibinfo {author} {\bibfnamefont {S.}~\bibnamefont
  {Caprara}}, \bibinfo {author} {\bibfnamefont {C.}~\bibnamefont {Di~Castro}},
  \bibinfo {author} {\bibfnamefont {G.}~\bibnamefont {Seibold}}, \ and\
  \bibinfo {author} {\bibfnamefont {M.}~\bibnamefont {Grilli}},\ }\href
  {\doibase 10.1103/PhysRevB.95.224511} {\bibfield  {journal} {\bibinfo
  {journal} {Phys. Rev. B}\ }\textbf {\bibinfo {volume} {95}},\ \bibinfo
  {pages} {224511} (\bibinfo {year} {2017})}\BibitemShut {NoStop}%
\bibitem [{\citenamefont {Berg}\ \emph {et~al.}(2012)\citenamefont {Berg},
  \citenamefont {Metlitski},\ and\ \citenamefont {Sachdev}}]{Berg12}%
  \BibitemOpen
  \bibfield  {author} {\bibinfo {author} {\bibfnamefont {E.}~\bibnamefont
  {Berg}}, \bibinfo {author} {\bibfnamefont {M.~A.}\ \bibnamefont {Metlitski}},
  \ and\ \bibinfo {author} {\bibfnamefont {S.}~\bibnamefont {Sachdev}},\ }\href
  {\doibase 10.1126/science.1227769} {\bibfield  {journal} {\bibinfo  {journal}
  {Science}\ }\textbf {\bibinfo {volume} {338}},\ \bibinfo {pages} {1606}
  (\bibinfo {year} {2012})}\BibitemShut {NoStop}%
\bibitem [{\citenamefont {Abanov}\ \emph {et~al.}(2003)\citenamefont {Abanov},
  \citenamefont {Chubukov},\ and\ \citenamefont {Schmalian}}]{Abanov03}%
  \BibitemOpen
  \bibfield  {author} {\bibinfo {author} {\bibfnamefont {A.}~\bibnamefont
  {Abanov}}, \bibinfo {author} {\bibfnamefont {A.~V.}\ \bibnamefont
  {Chubukov}}, \ and\ \bibinfo {author} {\bibfnamefont {J.}~\bibnamefont
  {Schmalian}},\ }\href {\doibase 10.1080/0001873021000057123} {\bibfield
  {journal} {\bibinfo  {journal} {Advances in Physics}\ }\textbf {\bibinfo
  {volume} {52}},\ \bibinfo {pages} {119} (\bibinfo {year} {2003})}\BibitemShut
  {NoStop}%
\bibitem [{\citenamefont {Schattner}\ \emph {et~al.}(2016)\citenamefont
  {Schattner}, \citenamefont {Gerlach}, \citenamefont {Trebst},\ and\
  \citenamefont {Berg}}]{Schattner16}%
  \BibitemOpen
  \bibfield  {author} {\bibinfo {author} {\bibfnamefont {Y.}~\bibnamefont
  {Schattner}}, \bibinfo {author} {\bibfnamefont {M.~H.}\ \bibnamefont
  {Gerlach}}, \bibinfo {author} {\bibfnamefont {S.}~\bibnamefont {Trebst}}, \
  and\ \bibinfo {author} {\bibfnamefont {E.}~\bibnamefont {Berg}},\ }\href
  {\doibase 10.1103/PhysRevLett.117.097002} {\bibfield  {journal} {\bibinfo
  {journal} {Phys. Rev. Lett.}\ }\textbf {\bibinfo {volume} {117}},\ \bibinfo
  {pages} {097002} (\bibinfo {year} {2016})}\BibitemShut {NoStop}%
\bibitem [{\citenamefont {Meier}\ \emph {et~al.}(2014)\citenamefont {Meier},
  \citenamefont {P\'epin}, \citenamefont {Einenkel},\ and\ \citenamefont
  {Efetov}}]{Pepin14}%
  \BibitemOpen
  \bibfield  {author} {\bibinfo {author} {\bibfnamefont {H.}~\bibnamefont
  {Meier}}, \bibinfo {author} {\bibfnamefont {C.}~\bibnamefont {P\'epin}},
  \bibinfo {author} {\bibfnamefont {M.}~\bibnamefont {Einenkel}}, \ and\
  \bibinfo {author} {\bibfnamefont {K.~B.}\ \bibnamefont {Efetov}},\ }\href
  {\doibase 10.1103/PhysRevB.89.195115} {\bibfield  {journal} {\bibinfo
  {journal} {Phys. Rev. B}\ }\textbf {\bibinfo {volume} {89}},\ \bibinfo
  {pages} {195115} (\bibinfo {year} {2014})}\BibitemShut {NoStop}%
\bibitem [{\citenamefont {Chowdhury}\ and\ \citenamefont
  {Sachdev}(2014{\natexlab{a}})}]{Chowdhury14}%
  \BibitemOpen
  \bibfield  {author} {\bibinfo {author} {\bibfnamefont {D.}~\bibnamefont
  {Chowdhury}}\ and\ \bibinfo {author} {\bibfnamefont {S.}~\bibnamefont
  {Sachdev}},\ }\href {\doibase 10.1103/PhysRevB.90.134516} {\bibfield
  {journal} {\bibinfo  {journal} {Phys. Rev. B}\ }\textbf {\bibinfo {volume}
  {90}},\ \bibinfo {pages} {134516} (\bibinfo {year}
  {2014}{\natexlab{a}})}\BibitemShut {NoStop}%
\bibitem [{\citenamefont {Li}\ \emph {et~al.}(2017)\citenamefont {Li},
  \citenamefont {Wang}, \citenamefont {Yao},\ and\ \citenamefont
  {Lee}}]{ZiXiang17}%
  \BibitemOpen
  \bibfield  {author} {\bibinfo {author} {\bibfnamefont {Z.-X.}\ \bibnamefont
  {Li}}, \bibinfo {author} {\bibfnamefont {F.}~\bibnamefont {Wang}}, \bibinfo
  {author} {\bibfnamefont {H.}~\bibnamefont {Yao}}, \ and\ \bibinfo {author}
  {\bibfnamefont {D.-H.}\ \bibnamefont {Lee}},\ }\href {\doibase
  10.1103/PhysRevB.95.214505} {\bibfield  {journal} {\bibinfo  {journal} {Phys.
  Rev. B}\ }\textbf {\bibinfo {volume} {95}},\ \bibinfo {pages} {214505}
  (\bibinfo {year} {2017})}\BibitemShut {NoStop}%
\bibitem [{\citenamefont {Chowdhury}\ and\ \citenamefont
  {Sachdev}(2014{\natexlab{b}})}]{Chowdhury14b}%
  \BibitemOpen
  \bibfield  {author} {\bibinfo {author} {\bibfnamefont {D.}~\bibnamefont
  {Chowdhury}}\ and\ \bibinfo {author} {\bibfnamefont {S.}~\bibnamefont
  {Sachdev}},\ }\href {\doibase 10.1103/PhysRevB.90.245136} {\bibfield
  {journal} {\bibinfo  {journal} {Phys. Rev. B}\ }\textbf {\bibinfo {volume}
  {90}},\ \bibinfo {pages} {245136} (\bibinfo {year}
  {2014}{\natexlab{b}})}\BibitemShut {NoStop}%
\bibitem [{\citenamefont {Atkinson}\ \emph {et~al.}(2015)\citenamefont
  {Atkinson}, \citenamefont {Kampf},\ and\ \citenamefont {Bulut}}]{Atkinson15}%
  \BibitemOpen
  \bibfield  {author} {\bibinfo {author} {\bibfnamefont {W.~A.}\ \bibnamefont
  {Atkinson}}, \bibinfo {author} {\bibfnamefont {A.~P.}\ \bibnamefont {Kampf}},
  \ and\ \bibinfo {author} {\bibfnamefont {S.}~\bibnamefont {Bulut}},\ }\href
  {http://stacks.iop.org/1367-2630/17/i=1/a=013025} {\bibfield  {journal}
  {\bibinfo  {journal} {New Journal of Physics}\ }\textbf {\bibinfo {volume}
  {17}},\ \bibinfo {pages} {013025} (\bibinfo {year} {2015})}\BibitemShut
  {NoStop}%
\bibitem [{\citenamefont {Wang}\ \emph {et~al.}(2017)\citenamefont {Wang},
  \citenamefont {Schattner}, \citenamefont {Berg},\ and\ \citenamefont
  {Fernandes}}]{Xiaoyu17}%
  \BibitemOpen
  \bibfield  {author} {\bibinfo {author} {\bibfnamefont {X.}~\bibnamefont
  {Wang}}, \bibinfo {author} {\bibfnamefont {Y.}~\bibnamefont {Schattner}},
  \bibinfo {author} {\bibfnamefont {E.}~\bibnamefont {Berg}}, \ and\ \bibinfo
  {author} {\bibfnamefont {R.~M.}\ \bibnamefont {Fernandes}},\ }\href {\doibase
  10.1103/PhysRevB.95.174520} {\bibfield  {journal} {\bibinfo  {journal} {Phys.
  Rev. B}\ }\textbf {\bibinfo {volume} {95}},\ \bibinfo {pages} {174520}
  (\bibinfo {year} {2017})}\BibitemShut {NoStop}%
\bibitem [{\citenamefont {Gerlach}\ \emph {et~al.}(2017)\citenamefont
  {Gerlach}, \citenamefont {Schattner}, \citenamefont {Berg},\ and\
  \citenamefont {Trebst}}]{Gerlach17}%
  \BibitemOpen
  \bibfield  {author} {\bibinfo {author} {\bibfnamefont {M.~H.}\ \bibnamefont
  {Gerlach}}, \bibinfo {author} {\bibfnamefont {Y.}~\bibnamefont {Schattner}},
  \bibinfo {author} {\bibfnamefont {E.}~\bibnamefont {Berg}}, \ and\ \bibinfo
  {author} {\bibfnamefont {S.}~\bibnamefont {Trebst}},\ }\href {\doibase
  10.1103/PhysRevB.95.035124} {\bibfield  {journal} {\bibinfo  {journal} {Phys.
  Rev. B}\ }\textbf {\bibinfo {volume} {95}},\ \bibinfo {pages} {035124}
  (\bibinfo {year} {2017})}\BibitemShut {NoStop}%
\bibitem [{\citenamefont {Wang}\ \emph {et~al.}(2015)\citenamefont {Wang},
  \citenamefont {Agterberg},\ and\ \citenamefont {Chubukov}}]{Yuxuan15}%
  \BibitemOpen
  \bibfield  {author} {\bibinfo {author} {\bibfnamefont {Y.}~\bibnamefont
  {Wang}}, \bibinfo {author} {\bibfnamefont {D.~F.}\ \bibnamefont {Agterberg}},
  \ and\ \bibinfo {author} {\bibfnamefont {A.}~\bibnamefont {Chubukov}},\
  }\href {\doibase 10.1103/PhysRevB.91.115103} {\bibfield  {journal} {\bibinfo
  {journal} {Phys. Rev. B}\ }\textbf {\bibinfo {volume} {91}},\ \bibinfo
  {pages} {115103} (\bibinfo {year} {2015})}\BibitemShut {NoStop}%
\bibitem [{\citenamefont {Moreo}\ and\ \citenamefont
  {Scalapino}(1991)}]{Moreo91}%
  \BibitemOpen
  \bibfield  {author} {\bibinfo {author} {\bibfnamefont {A.}~\bibnamefont
  {Moreo}}\ and\ \bibinfo {author} {\bibfnamefont {D.~J.}\ \bibnamefont
  {Scalapino}},\ }\href {\doibase 10.1103/PhysRevLett.66.946} {\bibfield
  {journal} {\bibinfo  {journal} {Phys. Rev. Lett.}\ }\textbf {\bibinfo
  {volume} {66}},\ \bibinfo {pages} {946} (\bibinfo {year} {1991})}\BibitemShut
  {NoStop}%
\bibitem [{Note1()}]{Note1}%
  \BibitemOpen
  \bibinfo {note} {Technically the hot-spots symmetry is SU(2)$\times $
  SU(2)$\sim $SO(4) and not SU(2), because for $\mu \not =0$ the CO order
  parameter is complex, giving rise to a four-component
  super-vector.}\BibitemShut {Stop}%
\end{thebibliography}
